\documentclass{iopart}

\usepackage{iopams}
\usepackage[dvips]{graphicx}

\newcommand{\bc}{\begin{center}}
\newcommand{\ec}{\end{center}}
\def\ba#1{\begin{array}{#1}\displaystyle}
\newcommand{\ea}{\end{array}}
\newcommand{\z}{\\ \displaystyle}

\newcommand{\beq}{\begin{equation}}
\newcommand{\eeq}{\end{equation}}
\newcommand{\beqa}{\begin{eqnarray}}
\newcommand{\eeqa}{\end{eqnarray}}
\newcommand{\no}{\nonumber}
\newcommand{\n}{\nonumber\\}
\def\mato#1{\left(\ba{#1}} 
\def\matf{\ea\right)}

\def\sect#1{\section{#1}\setcounter{equation}{0}}
\def\ssect#1{\subsection{#1}}
\def\sssect#1{\subsubsection{#1}}

\def\lt#1{\left#1}
\def\rt#1{\right#1}
\def\t#1{\tilde{#1}}

\def\b#1{\bar{#1}}
\def\frc#1#2{\frac{#1}{#2}}

\newcommand{\G}{\Gamma}
\newcommand{\p}{\partial}

\newcommand{\bra}{\langle}
\newcommand{\ket}{\rangle}
\newcommand{\Z}{{\mathbb{Z}}}
\newcommand{\N}{{\mathbb{N}}}

\newcommand{\Orr}{{\cal O}}

\newcommand{\varep}{\varepsilon}

\begin{document}

\title[Calogero-Sutherland eigenfunctions and CFT]{Calogero-Sutherland eigenfunctions with mixed
boundary conditions and conformal field theory correlators}

\author{B Doyon$^1$ and J Cardy$^{1,2}$}

\address{$^1$ Rudolf Peierls Centre for Theoretical
Physics, 1 Keble Road, Oxford OX1 3NP, UK}
\address{$^2$ All Souls College, Oxford}
\ead{b.doyon1@physics.ox.ac.uk}
\begin{abstract}
We construct certain eigenfunctions of the Calogero-Sutherland
hamiltonian for particles on a circle, with mixed boundary
conditions. That is, the behavior of the eigenfunction, as
neighbouring particles collide, depend on the pair of colliding
particles. This behavior is generically a linear combination of
two types of power laws, depending on the statistics of the
particles involved. For fixed ratio of each type at each pair of
neighboring particles, there is an eigenfunction, the ground
state, with lowest energy, and there is a discrete set of
eigenstates and eigenvalues, the excited states and the energies
above this ground state. We find the ground state and special
excited states along with their energies in a certain class of
mixed boundary conditions, interpreted as having pairs of
neighboring bosons and other particles being fermions. These
particular eigenfunctions are characterised by the fact that they
are in direct correspondence with correlation functions in
boundary conformal field theory. We expect that they have
applications to measures on certain configurations of curves in
the statistical $O(n)$ loop model. The derivation, although completely
independent from results of conformal field theory, uses ideas
from the ``Coulomb gas'' formulation.
\end{abstract}

\maketitle

\section{Introduction}

Recently \cite{Cardy04}, the Calogero-Sutherland quantum-mechanical hamiltonian (see, for instance, the book
\cite{Sutherland04}) was shown to be related to certain
bulk-boundary correlation functions in conformal field theory on
the disk. The Calogero-Sutherland hamiltonian
for $N$ particles at angles $\theta_1,\ldots,\theta_N$ on the
cirlce, with parameter $\beta$, is \beq\label{CSHamil}
    H_N(\beta) = -\sum_{j=1}^N \frc12\frc{\p^2}{\p\theta_j^2} +
    \frc{\beta(\beta-2)}{16} \sum_{1\leq j < k \leq N} \frc1{\sin^2\lt(\frc{\theta_j-\theta_k}2\rt)}~.
\eeq
The corresponding CFT has central charge $c$ related to the parameter $\beta$ through
\[
    \beta = \frc{8}{\kappa}~,\quad c = 1-\frc{3(4-\kappa)^2}{2\kappa}~.
\]
The hamiltonian is invariant under $\beta\to 2-\beta$. The
relations above imply that we chose the range $\beta\in[1,\infty]$
for the values $\kappa\in [0,8]$ that we will consider in this
paper. This discovery initially came from an analysis of the
equations believed to be associated to multiple SLE$_\kappa$
processes (Schramm-Loewner Evolution (SLE) processes were introduced in \cite{Schramm99},
multiple SLE generalisations were introduced in \cite{Cardy03} and
developed to a large extent in \cite{BauerBK05}, although only
through conjectured properties -- see \ref{defSLE} of a short review of what SLE is). But the connection can be
established solely from CFT concepts, as was shown in
\cite{Cardy04}. The main ingredients are level-2 degenerate
boundary fields, one for each particle, and a bulk primary field
at the center of the disk: the $N$ null-vector differential
operators \cite{BPZ84} acting on the correlation functions can be
recast, by taking a linear combination, into the
Calogero-Sutherland hamiltonian. Hence, the correlation functions
can then be recast into eigenfunctions of the hamiltonian. Various
choices of primary field give rise to various eigenfunctions; in
particular, the dimension of the bulk primary field is connected
to the energy associated to the eigenfunction. But not all
eigenfunctions can be reproduced in this way, since the $N$
null-vector equations are more restrictive than the one eigenvalue
equation of the hamiltonian. Two problems arise then naturally: to
determine which eigenfunctions (that is, which boundary
conditions, and for these boundary conditions, which states)
indeed give rise to correlation functions, and to obtain explicit
expressions for these eigenfunctions. These two problems are
solved in great part in this paper.

Finding eigenfunctions of the hamiltonian (\ref{CSHamil}) requires one more piece of information: the behavior of eigenfunctions $\Psi$ as particles collide (boundary conditions). Fixing the boundary conditions (which we will sometimes refer to as choosing a sector) fixes the Hilbert space; we will be more precise in the text about how boundary conditions are fixed. From the CFT
viewpoint, these behaviors are related to the boundary operator
product expansion (OPE) (more precisely, the overlap between the
bulk primary field and the boundary OPEs). Here and in the
following, we choose the sector
$\theta_1>\ldots>\theta_N>\theta_1-2\pi$, and we will consider the
behavior of eigenfunctions at the collisions $\theta_i\to
\theta_{i+1}^+$ (with $\theta_{N+1} \equiv \theta_1-2\pi$). It
will be sufficient to specify the behavior of an eigenfunction at
these boundaries in order to fix the eigenfunction\footnote{ The
behavior at collisions pertaining to other ordering of the angles
can in principle be obtained by analytic continuation.}. An
elementary indicial analysis of the Calogero-Sutherland system
shows that the behavior of the wave function as two particles
collide is generically a linear combination of two types of power
laws, which we will refer to as ``bosonic'': \beq\label{bbc}
    \Psi \propto (\theta_i-\theta_{i+1})^{\frc{\kappa-4}{\kappa}}\quad
    (\theta_i-\theta_{i+1}\to0^+)~;
\eeq and ``fermionic'': \beq\label{fbc}
    \Psi \propto (\theta_i-\theta_{i+1})^{\frc{4}{\kappa}}\quad (\theta_i-\theta_{i+1}\to0^+)~.
\eeq
This nomenclature comes from the fact that for $0<\kappa<4$ the wave function vanishes as
fermions (particles with fermionic bondary conditions) collide, whereas it diverges as bosons
(particles with purely bosonic boundary conditions) collide.
From the conformal field theory viewpoint, these correspond to the two families appearing in the
fusion of level-2 null fields: that of the identity, and that of level-3 null fields.
It is a simple matter to verify, for instance,
that the ground state in the sector with fermionic boundary conditions at all pairs of
colliding particles, which is the usual fermionic ground state,
do correspond to a correlation function satisfying
all null-vector equations, but the ground state with all bosonic boundary conditions generically does not.

In this paper we solve the null-vector equations for certain bulk
primary operators, of various scaling dimensions and of any spin.
The results give rise to integral formulas for certain
eigenfunctions of the Calogero-Sutherland hamiltonian. These
integral formulas are in close relation with those obtained by
Dub\'edat \cite{Dubedat05}, who was essentially considering the
case without bulk field. The technique we use is at the basis of
the Coulomb gas formalism \cite{DotsenkoF84, DotsenkoF85} of CFT
for bulk correlation functions in minimal models, and works for
generic central charge and for boundary operators as well. This
technique was also used in \cite{Kytola06} for correlation
functions without bulk field\footnote{We wish to note that the
paper \cite{Kytola06} was published after we had established the
working of our technique.}. The motivation was to evaluate the
``auxiliary functions'' appearing in constructions of multiple SLE
processes \cite{BauerBK05}, which satisfy the level-2 null-vector
equations with zero-dimension bulk field. In the present paper, we
will derive the formulas in the simplest way possible; we do not
need any of the machinery developed for the Coulomb gas formalism,
for CFT or for SLE, as we  work only with the differential
equations.

The construction gives rise to a certain class of boundary
conditions for the eigenfunctions, satisfied by the ground states
and the excited states. It is important to understand that with
mixed boundary conditions, one should only distinguish between
classes $C$ whose elements can be obtained from one another by
simple linear combinations
\[
    \Psi_1 \in C ~\mbox{and}~ \Psi_2\in C \Rightarrow \Psi_3 = a \Psi_1 + b\Psi_2 \in C ~\mbox{for}~ a\ge 0, b\ge0~.
\]
If $\Psi_1$ and $\Psi_2$ have different mixed boundary conditions
but correspond to the same energy, than $\Psi_3$ also is an
eigenfunction of the hamiltonian, with yet again different mixed
boundary conditions and with the same energy. Also, if both
$\Psi_1$ and $\Psi_2$ are ground states, everywhere positive, then
$\Psi_3$ also is (this is why we need the condition that both $a$
and $b$ be greater than zero: the eigenfunction of a ground state
should be everywhere positive). It is easy to obtain, from
$\Psi_1$ with a given energy, $\Psi_2$ with the same energy: one
only needs, for instance, to make cyclic permutations of the
particle positions.

\begin{figure}[t]
\bc \includegraphics{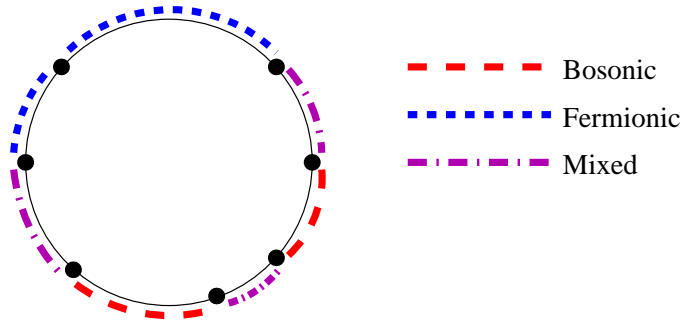} \ec
\caption{An example of boundary conditions satisfied by our solutions.}
\label{figintrobound}
\end{figure}

The classes of boundary conditions that we obtained are those with
distinctive elements as follows: some pairs of colliding
neighouring particles, which do not have
common members among each other, present purely bosonic behavior (that is,
the eigenfunction behaves like (\ref{bbc}) times a power series in
$(\theta_i-\theta_{i+1})$), pairs formed by any other neighboring particles present purely fermionic behavior,
and the remaining pairs present both fermionic and bosonic components, in a
certain fixed proportion (see Fig. \ref{figintrobound}). One can
interpret the pairs with purely bosonic behavior as being pairs of
bosons, whereas the other particles as being fermions. Note that
in one dimension this does not have any implication for the way
the eigenfunction should behave when non-neighboring particles
approach each other: we work only with a fixed ordering of the
$\{\theta_j\}$. While the eigenfunction may be analytically
continued to other orderings, these are not physical. This is in
distinction to the case in higher dimensions, when particles can
be moved past each other.

The Calogero-Sutherland system was mainly studied, until now, on
the Hilbert space of wave functions with simple uniform boundary
conditions: all particle collisions giving only fermionic
exponents, or all giving only bosonic exponents. Our new solutions
to the null-vector equations give special eigenfunctions of the
Calogero-Sutherland system for mixed boundary conditions,
physically corresponding to some particles being
fermions and some being bosons. An eigenfunction of
the Calogero-Sutherland hamiltonian with non-zero (angular)
momentum, which would correspond to a bulk primary field with
non-zero spin, can always be obtained from one with zero momentum
by a Galilean transformation. However, such eigenfunctions are not
generically in agreement with all null-vector equations. Our
solutions with non-zero spin are not simple Galilean transform of
those with zero spin. They are yet new solutions, and correspond, in fact, to giving non-zero
momentum only to the fermions.

\begin{figure}[t]
\bc \includegraphics{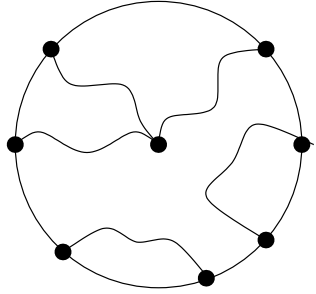} \ec
\caption{An example of a domain-wall configuration in the $O(n)$ model, or multiple SLE configuration,
corresponding to our solution with the 3-leg operator.}
\label{figintroconf}
\end{figure}

Some of the bulk primary operators corresponding to our solutions
are the ``$N'$-leg operators'' (with $N'=N-2M,\,M\in\N$) -- that is, their dimensions
are the ``$N'$-leg'' exponents \cite{SaleurD87}. They have meaning in the context of the critical $O(n)$ loop model
\cite{Nienhuis82}, and they are expected to be connected to certain restriction of or events in
multiple-SLE measures. The corresponding solutions are expected to be related to measures
for configurations of the type shown in Fig. \ref{figintroconf} (although we could only give a conjecture for this relation in the case
where there is single pairing).
It is these solutions that lead to ground states. Other operators
correspond to excited states and to states with non-zero momentum,
but have as yet no known physical interpretation.

The paper is organised as follows. In Section \ref{review}, we recall the results of \cite{Cardy04}. Then,
in Section \ref{sectintegrep}, we construct our integral representations, and derive the associated
boundary conditions. We also give the Coulomb-gas interpretation of our construction. In Section \ref{N3},
we derive some general results about solutions to the null-vector equations in the case where $N=3$
(we show that our solutions form a complete basis). Finally, in Section \ref{sectDiscuss}, we discuss
the interpretation of our results in the continuum $O(n)$ loop model.

\sect{Review of the connection between null vector equations and the Calogero-Sutherland system}
\label{review}

\ssect{Null vector equations}

Consider the following family of correlation functions in a boundary conformal field theory on the unit disk:
\beq\label{corrfct}
    G = \bra \phi(e^{i\theta_1})\cdots \phi
    (e^{i\theta_N})\,\Phi(0)\ket
\eeq where $\phi$ are primary level-2 degenerate boundary fields
and $\Phi$ is a primary bulk field, for $N=1,2,\ldots$. For now we
will consider only spinless bulk fields $\Phi$, deferring the
discussion of a field with spin to Section \ref{sectspin}. Using
the parameter $\kappa$ appearing naturally in SLE, we will
parametrize the central charge $c$ and the dimension $h$ of the
boundary fields by \beq\label{paramch}
    c = 1-\frc{3(4-\kappa)^2}{2\kappa}~,\quad
    h=\frc{6-\kappa}{2\kappa}~.
\eeq As was shown in \cite{Cardy04}, the null-vector equations
\cite{BPZ84,DiFrancescoMathieuSenechal97} associated to this
correlation function imply that a certain simple modification of
this correlation function is an eigenfunction of the
Calogero-Sutherland hamiltonian. One applies the infinitesimal
conformal transformation $z\mapsto z + \alpha(z)$ with
\beq\label{alpha}
    \alpha(z) = \sum_{j=1}^N b_j\alpha_j(z)~,\quad
    \alpha_j(z) = -z\frc{z+e^{i\theta_j}}{z-e^{i\theta_j}}~.
\eeq
Thanks to the relation
\beq\label{presD}
    \overline{\alpha_j(z)} = -\bar{z}^{2}
    \alpha_j(\bar{z}^{-1})~,
\eeq
this infinitesimal transformation preserves the region $\mathbb{D} \setminus \{z_j\}$,
the disk minus the boundary points $z_j \equiv e^{i\theta_j}$. It is a pure scaling at the center
$\alpha(z)\sim z,\,z\to0$, and it has poles at the positions $z_j$ of the boundary fields, generating locally
there non-trivial conformal transformations whose effect can be evaluated thanks to the null-vector property.
The result is the set of differential equations
\beq\label{nveq}
    \sum_{j=1}^N b_j{\cal D}_j G  =  d_\Phi \lt(\sum_{j=1}^N b_j\rt) G
\eeq
where $d_\Phi$ is the scaling dimension of $\Phi$, and with
\beqa\label{Dj}
    {\cal D}_j &=& -\frc{\kappa}2 \lt(\frc{\p}{\p \theta_j}\rt)^2
    + \frc{(6-\kappa)(\kappa-2)}{8\kappa} -\\ &&
    - \sum_{k\neq
    j}\lt(\cot\lt(\frc{\theta_k-\theta_j}2\rt)\frc{\p}{\p\theta_k}
    - \frc{h}{2\sin^2\lt(\frc{\theta_k-\theta_j}2\rt)}\rt)~.
\eeqa
This is derived in \ref{derivDj} for completeness of the discussion. Note further that
\[
    \frc{(6-\kappa)(\kappa-2)}{8\kappa} = \frc{h}6 + \frc{c}{12}~.
\]
This reproduces the operator ${\cal D}_j$ obtained in \cite{Cardy04} from slightly different arguments.

\ssect{Calogero-Sutherland hamiltonian}

It will be convenient for this paper to introduce the notation
\beq
    f(\theta) = \cot\lt(\frc\theta2\rt)~,\quad f_{jk} =
    f(\theta_j-\theta_k)~,\quad F_j = \sum_{k\neq j} f_{jk}~.
\eeq
Using this notation, we have
\beq
    {\cal D}_j = -\frc{\kappa}2 \p_j^2
    + \frc{(6-\kappa)(\kappa-2)}{8\kappa}
     + \sum_{k\neq
    j}\lt(f_{jk} \p_k - h f_{jk}'\rt)
\eeq with $\p_j \equiv \p / \p\theta_j$, and the
Calogero-Sutherland hamiltonian (\ref{CSHamil}) can be written
\beq
    H_N(\beta) = -\sum_j \lt(\frc12\p_j^2 + \frc{\beta(\beta-2)}{16} F_j' \rt)~.
\eeq
In order to relate the null-vector equations to the Calogero-Sutherland system, we look at the case
where $b_j=1$ for $j=1,\ldots,N$. We will denote
\beq
    {\cal D} = \sum_j {\cal D}_j = -\frc{\kappa}2 \sum_j\p_j^2
    + N\frc{(6-\kappa)(\kappa-2)}{8\kappa} \n
     - \sum_{j}\lt(F_j \p_j + h F_j'\rt)~.
\eeq
Equation (\ref{nveq}) implies that correlation functions $G$ are eigenfunctions of ${\cal D}$ with
eigenvalue $Nd_\Phi$. Consider the function of all $\theta_j$'s
\beq\ba{l}
    g_r = \prod_{1\leq j< k\leq N} \lt(
    \sin\frc{\theta_j-\theta_k}2\rt)^{-2r}\z
    (\theta_i>\theta_{i+1},\;i=1,\ldots,N-1;\; \theta_{N} > \theta_1-2\pi)
    \ea
\eeq
From the properties $\p_j g_r = -r\,g_rF_j$ and $\sum_j F_j^2 = -2 \sum_j F_j' - \frc{N(N^2-1)}3$,
it is a simple matter to check that
\beq
    g_{-\frc1\kappa} \cdot {\cal D} \cdot g_{\frc1\kappa} =
        \kappa H_N\lt(\frc{8}{\kappa}\rt)
        -\frc{N(N^2-1)}{6\kappa}
        + N\frc{(6-\kappa)(\kappa-2)}{8\kappa}
\eeq
(here and below, the dot ($\cdot$) means multiplication as operators on functions).
Hence, any correlation function $G$ gives rise to an eigenfunction
\beq\label{relEigen}
    \Psi = g_{\frc1{\kappa}}^{-1} G
\eeq
of the Calogero-Sutherland hamiltonian
$H_N\lt(\frc{8}{\kappa}\rt)$, with eigenvalue \beq\label{relEd}
    E = \frc{N}{\kappa} \lt[d_\Phi +\frc{(N^2-1)}{6\kappa}
        - \frc{(6-\kappa)(\kappa-2)}{8\kappa}\rt]~.
\eeq

\ssect{The fermionic and bosonic ground states}

The set of null-vector equations (\ref{nveq}) is more restrictive
than the eigenvalue equations for the hamiltonian (\ref{CSHamil}).
Hence, not all eigenfunctions satisfy all requirements to be
associated to CFT correlation functions. Here we recall the
fermionic and bosonic ground states of the Calogero-Sutherland
hamiltonian, and verify in which case they can be associated to
correlation functions.

It is a simple matter to find certain eigenfunctions of the operator ${\cal D}$, which correspond
to the ground state of $H_N\lt(\frc{8}{\kappa}\rt)$ with all fermionic (\ref{fbc}) or
all bosonic (\ref{bbc}) boundary conditions. Indeed, we have
\beqa
    g_{-r} \cdot {\cal D} \cdot g_{r}\no
    &=& -\frc{\kappa}2 \sum_j\p_j^2  +
     (\kappa r-1)\sum_jF_j\p_j + \n && + \lt(-2r+\kappa r^2 + \frc{\kappa
    r}2 - h\rt)\sum_j F_j' + \n &&
    + N\frc{(6-\kappa)(\kappa-2)}{8\kappa} + \lt(-r+\frc{\kappa r^2}2\rt)
    \frc{N(N^2-1)}3\no
\eeqa
so that with the two values
\beq
    r=r_f\equiv -\frc1{\kappa} ~,\quad r=r_b \equiv h =
    \frc{6-\kappa}{2\kappa}
\eeq the factor multiplying $\sum_jF_j'$ vanishes (note that these
two values are equal, $r_f=r_b$, only at $\kappa=8$). Hence, a
simple eigenfunction of this operator is 1, which gives the usual
fermionic and bosonic ground-state eigenfunctions of the
Calogero-Sutherland hamiltonian (see for instance \cite{Sutherland04}) \beq\label{gswavefct}
    \Psi^f_N = g_{r_f-\frc1\kappa} = g_{-\frc{2}\kappa}~,\quad
    \Psi^b_N = g_{r_b-\frc1\kappa} = g_{\frc{4-\kappa}{2\kappa}}
\eeq
with the associated eigenvalues
\beq
    E^f_N =\lt(\frc{4}\kappa\rt)^2 \frc{N(N^2-1)}{24}~,\quad
    E^b_N =\lt(\frc{4-\kappa}\kappa\rt)^2 \frc{N(N^2-1)}{24}~.
\eeq They are related by the transformation $4/\kappa\to
1-4/\kappa$ keeping the hamiltonian $H_N(8/\kappa)$ invariant.

The corresponding eigenfunctions of the operator $\cal D$ are
\beq\label{Nlegcorrfct}
    G^f_N = g_{r_f} ~,\quad G^b_N = g_{r_b}
\eeq
and the eigenvalues are $Nd^f_N, Nd^b_N$ with
\beq\label{Nlegexp}
    d^f_N = \frc{N^2}{2\kappa} - \frc{(\kappa-4)^2}{8\kappa}
\eeq
and
\beq
    d^b_N = \frc{(6-\kappa)(\kappa-2)}{24\kappa}(4 -
    N^2)~.
\eeq

Note that in general, any correlation function $G$ with behavior $\propto (\theta_i-\theta_{i+1})^{-2r_f}$
leads to the fermionic boundary condition (\ref{fbc}) for the wave function, whereas
any correlation function with behavior $\propto (\theta_i-\theta_{i+1})^{-2r_b}$
leads to the bosonic boundary condition (\ref{bbc}).

It is a simple matter to check that $G^f_N$ satisfies all equations (\ref{nveq}),
but that $G^b_N$ does not, unless $N=2$ or $\kappa=6$ or $\kappa=8$ (in the latter case, $G^b_N = G^f_N$).
Hence, $G^f_N$ is a CFT correlation function,
and $d^f_N$ is a scaling dimension of a primary operator that couples to boundary level-two null vectors;
whereas $G^b_N$ and $d^b_N$ generically are not. Note that $d^f_N$ is equal to the $N$-leg
exponent \cite{SaleurD87}. Consider the similarity transform
\beqa
    g_{-r} \cdot {\cal D}_j \cdot g_r &=& -\frc{\kappa}2 \p_j^2 + \kappa r
    F_j\p_j - \frc{\kappa r^2}2 F_j^2 + \frc{\kappa r}2 F_j' + \n && +
     \sum_{k\neq j} \lt(f_{jk}(\p_k - rF_k) - hf_{jk}'\rt) +
     \frc{(6-\kappa)(\kappa-2)}{8\kappa}~.
\eeqa
The function $g_r$ is an eigenfunction of ${\cal D}_j$ if and only if the term
\[
    - \frc{\kappa r^2}2 F_j^2 + \lt(\frc{\kappa r}2-h\rt) F_j'
    -r \sum_{k\neq j} f_{jk} F_k  +
     \frc{(6-\kappa)(\kappa-2)}{8\kappa}
\]
is a constant. Some algebra (or a simple analysis of the simple and double poles)
shows that it is indeed constant if and only if
\beq
    -2\kappa r^2-\kappa r + 2h +4r =0 \quad \mbox{and} \quad r(2\kappa r+2) \sum_{k\neq j,\,k\neq l} f_{lk} =0
\eeq
for all $l\neq j$. The first condition is satisfied for $r=r_f$ or $r=r_b$ only, and the second,
for $N=2$ or $r=r_f$ or $r=0$. Hence, for $N>2$, $G^f_N$ is a common
eigenfunction of all ${\cal D}_j$ for any $\kappa$, and $G^b_N$ is
only for $\kappa=6$ (making $h=0$) or $\kappa=8$ (in which case $r_f=r_b$). In the case $\kappa=6$, the
function $G^b_N$ is just a constant. One can also check that the eigenvalues are independent of $j$:
\beq
    {\cal D}_j g_r = \lt(\frc{\kappa r^2}2(N-1)^2 - r
    (N-1) + \frc{(6-\kappa)(\kappa-2)}{8\kappa}\rt) g_r
\eeq
in the cases above. The eigenvalues are indeed equal to $d_{\Phi_N^{f}}$ and $d_{\Phi_N^{b}}$
when we put, repectively, $r=r_f$ and $r=r_b$.

\ssect{$L^2$-normalisability and hermiticity}

The fermionic ground-state eigenfunction $\Psi_f$ (\ref{gswavefct}) is $L^2$-normalisable for the full range $0<\kappa\leq 8$, but the bosonic one, $\Psi_b$, is only for $8/3<\kappa\leq 8$ (the value $\kappa=8/3$ is the value at wich the bosonic behaviour is of power $-1/2$ in wave functions). In general, as soon as a wave function has bosonic behaviour at some colliding pair of angles, it is $L^2$-normalisable only in that range; in particular, this holds for all wave functions in sectors with mixed boundary conditions found below. For generic $\kappa$ in the normalisable range, the hamiltonian (\ref{CSHamil}) is hermitian. This is easy to understand from the $L^2$ norm. With $\Psi_1$ and $\Psi_2$ hamiltonian eigenfunctions (possibly in a mixed sector), one evaluates $\int_\varepsilon d\theta_1\cdots d\theta_N \,\Psi_1^* H\Psi_2$ where the integration region is $\theta_i>\theta_{i+1}+\varepsilon$ (and $\theta_{N+1} \equiv \theta_1-2\pi$) for some small positive $\varepsilon$. By normalisability and by the eigenfunction property, this multiple integral converges as $\varepsilon\to0$. Checking hermiticity involves integrating by part on all angles (to make things more obvious, one could change variables to angle differences and the total angle average), and the only possible violation of hermiticity comes from boundary terms as angle differences are equal to $\varepsilon$. But for generic $\kappa$, these will be non-integer (possibly negative) powers of $\varepsilon$, generically not the power 0. Since the multiple integral resulting after integration by part is also convergent, all boundary terms must vanish as $\varepsilon\to0$, which shows hermiticity.

\sect{Integral representations of solutions to null-vector equations: mixed boundary
conditions and excited states} \label{sectintegrep}

In this section, we construct integral representations for solutions to the null-vector equations (\ref{nveq})
employing a technique that mimics the Coulomb gas formalism of CFT. We will observe that some of these
solutions correspond to excited states of the Calogero-Sutherland system above the completely fermionic
ground state, that some correspond to completely new ground-state solutions with
boundary conditions that are purely bosonic at certain pairs of colliding angles and purely fermionic at other
pairs (as described in the introduction), and that some are excited states above these new ground states.
The results of this section are very similar in form to those of Dub\'edat \cite{Dubedat05}, and the techniques
are in close relations to those used in \cite{Kytola06}.

\ssect{One integration variable}

Consider the function
\beq\label{w}
    w = G^f_N \prod_{1\leq k\leq N}
    \lt|\sin\frc{\theta_k-\zeta}2\rt|^{-2\alpha}~.
\eeq
Denote
\beq
    f_j = f(\theta_j-\zeta)~.
\eeq
Then, we have
\beq
    \p_j w = \frc{F_j}{\kappa} w - \alpha f_j w~.
\eeq
Consider also the new operator
\beq\label{Wj1}
    {\cal W}_j = {\cal D}_j + f_j\p_\zeta - f_j'~.
\eeq
One finds that
\beqa
    (w^{-1} \cdot {\cal W}_j \cdot w)1 &=&
    \frc{(6-\kappa)(\kappa-2)}{8\kappa} +
    \frc{N^2-1}{2\kappa} + \lt(\frc{\kappa\alpha}2 -1\rt)  f_j' - \n && -
    \alpha\lt(\frc{\kappa\alpha}2 -1\rt) f_j^2
    - \alpha\sum_{k\neq j} (f_{jk}(f_k-f_j) - f_kf_j)~.
\eeqa
In order to cancel the double pole at $\theta_j=\zeta$ coming from $f_j'$ and $f_j^2$, we need
\beq
    \alpha\lt(\frc{\kappa\alpha}2 - 1\rt) = -\frc12
    \lt(\frc{\kappa\alpha}2 -1\rt)
\eeq
so that
\beq\label{valalpha}
    \alpha = -\frc12 \qquad \mbox{or} \qquad \alpha =
    \frc2{\kappa}~.
\eeq
It is a simple matter to verify that the sum $\sum_{k\neq j}
(f_{jk}(f_k-f_j) - f_kf_j)$ does not have poles at
$\theta_j=\zeta$, $\theta_j=\theta_k\;(k\neq j)$ and
$\theta_k=\zeta \; (k\neq j)$, so that it is a constant.
Evaluating this constant by taking $\theta_j\to-i \infty$,
where $f_{jk} = f_j = i$, we find:
\beq
    {\cal W}_j  w = \lt[\frc{N^2}{2\kappa} - \frc{(\kappa-4)^2}{8\kappa} + \alpha\lt(\frc{\kappa\alpha}2 -
    N\rt)\rt] w \equiv d^{(\alpha)}_N w~.
\eeq
In the first case of (\ref{valalpha}), the eigenvalue is given by
\beq\label{EigenWj1e}
    d^{\lt(-\frc12\rt)}_N = \frc{(N+\kappa/2)^2}{2\kappa} -
    \frc{(\kappa-4)^2}{8\kappa}
\eeq
whereas in the second case, it is
\beq\label{EigenWj1}
       d^{\lt(\frc2\kappa\rt)}_N= \frc{(N-2)^2}{2\kappa} - \frc{(\kappa-4)^2}{8\kappa}
       = d^f_{N-2}~.
\eeq We now consider the analytic continuation of $w$ as function
of $\zeta$. For definiteness, we choose the analytic
continuation from the region $\theta_N>\zeta>\theta_1 - 2\pi$,
where it is real and positive, and we still denote this analytic
continuation by $w$. Note that \beq
    {\cal W}_j = {\cal D}_j + \p_\zeta \cdot f_j~.
\eeq
Hence, the function
\beq\label{integexpr1}
    G_{\cal C} = A \oint_{\cal C} d\zeta \; w
\eeq
satisfies
\beq
    {\cal D}_j G_{\cal C} = d^{(\alpha)}_N G_{\cal C}
\eeq
for any closed contour ${\cal C}$ on the multi-sheeted Riemann surface on which $w$ lives as a function of $\zeta$;
the function $G_{\cal C}$ will be non-zero only for contours that are topologically non-trivial. The
normalisation constant $A$ will be chosen for convenience: if possible, it will be such that the result is
real and positive in the chosen sector $\theta_1>\cdots>\theta_N>\theta_1-2\pi$. This
is necessary for identifying the result as a ground state of the Calogero-Sutherland system (that is, without
zeros), as well as for its interpretation as a measure on stochastic processes (but obviously not necessary
for the interpretation as correlation functions, or as linear combinations of measures with complex coefficients).

In fact, the analytic structure of the integration measure
$d\zeta\, w$ is easier to see when it is expressed in terms of the
variables $z_j = e^{i\theta_j}$ and $y = e^{i\zeta}$. In terms of
these variables, the function $w$ is \beq
    w = G^f_N (2i)^{2\alpha N} y^{\alpha N} \prod_{1\leq k\leq N} \lt[z_k^{\alpha} (z_k-y)^{-2\alpha}\rt]~.
\eeq The analytic structure of the function $d\zeta/dy\;w=-i
\,w/y$ is as follows:
\begin{itemize}
\item Case $\alpha = -\frc12$. There are two singular points: one at $y=0$, of the type $y^{-1-N/2}[[y]]$,
and one at $y=\infty$, of the type $y^{-1+N/2}[[y^{-1}]]$.
\item Case $\alpha = \frc2\kappa$. There are singular points at $y=z_j$ of the type $(y-z_j)^{-4/\kappa}[[y-z_j]]$,
at $y=0$ of the type $y^{-1+2N/\kappa}[[y]]$ and at $y=\infty$ of the type $y^{-1-2N/\kappa}[[y^{-1}]]$.
\end{itemize}

\sssect{Case $\alpha=-\frc12$: excited state above the fermionic ground state}

\begin{figure}[t]
\bc \includegraphics{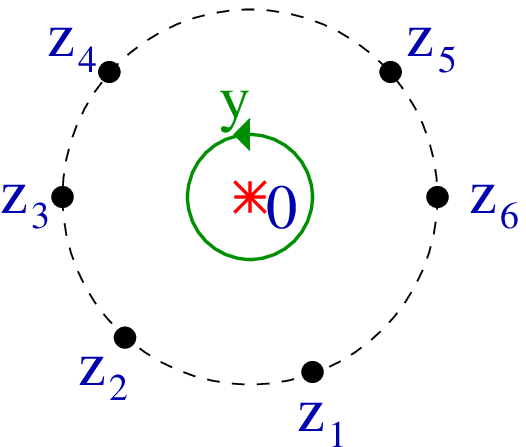} \ec
\caption{Contour ${\cal C}_{\rm origin}$ (with $N=6$).}
\label{contour-origin}
\end{figure}

If $N$ is even, there is only one class of topologically non-trivial contours, those circling the origin.
Circling once counterclockwise (contour ${\cal C}_{\rm origin}$, see Fig. \ref{contour-origin}),
the result is (with appropriate normalisation)
\beqa
    G_{{\cal C}_{\rm origin}} &=&
    G_N^f \sum_{u\equiv\{u_1,\ldots,u_{N/2}\},v\equiv\{v_1,\ldots,v_{N/2}\}\atop
      |\ u \cup v = \{1,\ldots,N\}}
    \cos\sum_{j=1}^{N/2} \frc{\theta_{u_j}-\theta_{v_j}}2 \n
    &=& \frc1{(N/2)!}\; G_N^f \sum_{\{\{u_1,v_1\},\ldots,\{u_{N/2},v_{N/2}\}\}\atop
      |\ \cup_{j=1}^{N/2} \{u_j,v_j\} =\{1,\ldots,N\}}
    \prod_{j=1}^{N/2} \cos\frc{\theta_{u_j}-\theta_{v_j}}2~.
    \label{G12}
\eeqa
Observe that although the result is clearly real, it is impossible to make it positive everywhere
(for any one ordering of the angles).

If $N$ is odd, there is only one class again, a representative
being the figure-8 contour circling the origin counterclockwise
and the point $\infty$ clockwise (or circling twice the origin
counterclockwise). However, because of the structure of the
singularities at the origin and at infinity, this gives zero:
there are no contours giving non-zero eigenfunctions with
eigenvalue $d_N^{\frc12}$ for $N$ odd.

The dimension $d_N^{\lt(-\frc12\rt)}$ associated to the
correlation function $G$ (\ref{G12}) corresponds to the energy of
a certain excited state of the Calogero-Sutherland hamiltonian
above the $N$-particle completely fermionic ground state
$\Psi_N^f$ (\ref{gswavefct}) for $N$ even. In general, these
excited states are characterised by a set of non-negative integers
$p_j,\,j=1,\ldots,N-1$, and have eigenvalues
$E_N^{f;\,p_1,\ldots,p_{N-1}} = \frc12 \sum_{j=1}^N k_j^2$ with
$\sum_{j=1}^N k_j = 0,~ k_{j+1}-k_j = 4/\kappa + p_j$ \cite{Sutherland04}. The
corresponding field dimension, related to the eigenvalues
$E_N^{f;\,p_1,\ldots,p_{N-1}}$ through (\ref{relEd}), will be
denoted $d_N^{f;\,p_1,\ldots,p_{N-1}}$. A configuration of $p_1,\ldots,p_{N-1}$ that
reproduces the dimension $d_N^{\lt(-\frc12\rt)}$ satisfies
\beqa
	\sum_{l=1}^{N-1} \sum_{l'=1}^{N-1} (N {\rm min}(l,l') - ll') p_l p_{l'} &=& \frc{N^2}4 \n
	\sum_{l=1}^{N-1} l(N-l) p_l = \frc{N^2}4~.\no
\eeqa
It is a simple matter to observe that there are no solutions for odd $N$, and that for all $N$ even,
\beq\label{excfermionic}\ba{l}
    d^{\lt(-\frc12\rt)}_N = d_N^{f;\,p_1,\ldots,p_{N-1}} \z \mbox{with} \quad \lt\{\ba{l}
    p_1=\cdots = p_{N/2-1} = p_{N/2+1} = \cdots = p_{N-1} = 0~, \\ p_{N/2}=1 \ea\rt. \z (N\ \mbox{even})\ea~.
\eeq
Also, one can check that there are no other configurations of $p_1,\ldots,p_{N-1}$ that reproduce $d^{\lt(-\frc12\rt)}_N$ for all even $N\leq 10$ (that is, these states are non-degenerate).

It is easy to check explicitly that our solution (\ref{G12}) for $N$ even, along with the transformation (\ref{relEigen}), reproduces the well-known eigenfunctions for these excited states. Indeed, the function multiplying $G_N^f$ is proportional to the sum over all $z_i$-permuttations of the product $\prod_i z_i^{\lambda_i}$ where half of the $\lambda_i$'s are $+1/2$, and half are $-1/2$. This corresponds to a single gap in the ``Fermi sea'' of particles, making it two filled bands with the same number of particles separated by the minimum energy, as described by the configuration $p_{N/2}=1$ and $p_j=0,\;j\neq N/2$, along with the constraint of zero total momentum (see, for instance, \cite{Sutherland04}).

Hence we have found that certain fermionic excited states of the
Calogero-Sutherland hamiltonian are in fact also solutions to all
null-vector equations (\ref{nveq}).

\sssect{Case $\alpha = \frc2\kappa$: ground state with mixed boundary conditions}

\begin{figure}[t]
\bc \includegraphics{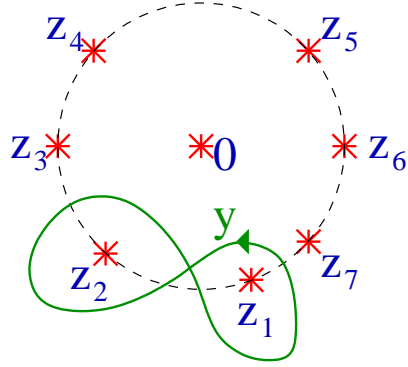} \ec
\caption{Contour ${\cal C}^{(1)}$ (with $N=7$).}
\label{contour-8}
\end{figure}

In this case, many classes of non-trivial contours exist. It turns
out that a basis can be obtained by taking figure-8 contours that
surround the point $z_i$ once counterclockwise and the point
$z_{i+1}$ once clockwise, for $i=1,\ldots,N$ (with $z_{N+1} \equiv
z_1$). We will denote contours of this type by ${\cal C}^{(i)}$
where $i$ stands for the index of the first member $z_i$ (in the
clockwise ordering around the unit circle) of a pair of adjacent
angles (see Fig. \ref{contour-8}). The result of the integration with any other
contour can be written as a linear combination of the integration
with the contours ${\cal C}^{(i)}$. Note that figure-8 contours
are closed since the singularities at the points $z_1,\ldots,z_N$
are all of the same type. The solutions that we consider are then
\beq\label{Gint}
    G_{{\cal C}^{(i)}} = A\;(2i)^{\frc{4N}\kappa} \; G_N^f\;  \int_{{\cal C}^{(i)}} dy\; y^{\frc{2N}\kappa -1}
     \prod_{1\leq j\leq N}\lt(z_j^{\frc{2}{\kappa}} (z_j-y)^{-\frc4\kappa} \rt) ~.
\eeq
The two parts of the integration contour that lie between $z_i$ and $z_{i+1}$ can be collapsed to a segment of line
(on different Riemann sheets), and if $\kappa>4$, the contributions around the points $z_i$ and $z_{i+1}$
can be set to zero by collapsing them upon the points $z_i$ and $z_{i+1}$, respectively. One is then left with
\beqa\label{Gintg4}
    G_{{\cal C}^{(i)}} &=&
    A\;(2i)^{\frc{4N}\kappa} \;(1- \omega)\; G_N^f\;\times\\ && \times \int_{z_i}^{z_{i+1}} dy\; y^{\frc{2N}\kappa -1}
    \prod_{1\leq j\leq N}\lt(z_j^{\frc{2}{\kappa}} (z_j-y)^{-\frc4\kappa} \rt) \n && (\kappa>4)\no
\eeqa
where
\[
    \omega = e^{\frc{8i\pi}\kappa}~.
\]
For the rest of this sub-section, we will restrict ourselve to the case $\kappa>4$.

This solution corresponds to the ground state of the Calogero-Sutherland system with certain mixed boundary
condition. That it is a ground state (that it has no zeros in the sector that we consider) is seen
by writing the expression (\ref{Gintg4}), with an appropriate normalisation constant,
in a form that is obviously real and positive ($A=1/2\;e^{i\varphi}$ for some real $\varphi$ that depends on $i$):
\beq\label{Gintg4real}
    (\kappa>4) \quad G_{{\cal C}^{(i)}} = \sin\frc{4\pi}\kappa\;
    G^f_N~\int_{\theta_i}^{\theta_{i+1}} d\zeta\; \prod_{1\leq k\leq N}
    \lt|\sin\frc{\theta_k-\zeta}2\rt|^{-\frc4\kappa}~.
\eeq
Below we show that it has purely bosonic behavior as $\theta_i\to \theta_{i+1}$, mixed
behavior as $\theta_{i-1}\to\theta_i$ and $\theta_{i+1}\to\theta_{i+2}$, and purely fermionic elsewhere.
That is, the solution takes the following forms when expanded around different pairs of colliding angles:
\beq\label{formbc1}\ba{l}
    G_{{\cal C}^{(i)}} = \z \lt\{\ba{l} (\theta_i-\theta_{i+1})^{-2r_b} [[\theta_i-\theta_{i+1}]] \\
       (\theta_{i+1}-\theta_{i+2})^{-2r_b} [[\theta_i-\theta_{i+1}]]
       +  (\theta_{i+1}-\theta_{i+2})^{-2r_f} [[\theta_{i+1}-\theta_{i+2}]] \\
       (\theta_{i-1}-\theta_{i})^{-2r_b} [[\theta_{i-1}-\theta_{i}]]
       +  (\theta_{i-1}-\theta_{i})^{-2r_f} [[\theta_{i-1}-\theta_{i}]] \\
       (\theta_{j}-\theta_{j+1})^{-2r_f} [[\theta_j-\theta_{j+1}]]
       \ea \rt.\ea
\eeq
where $j\neq i,i+1,i-1 \ {\rm mod} \ N$.

Let us analyse the boundary conditions from the expression (\ref{Gintg4}).
Take for simplicity $i=1$ (other cases are obtained by a cyclic permuttation of the variables).
The singularity as $z_1\to z_2$ can be obtained by setting the variables $y$ and $z_1$ to $z_2$ everywhere
except in the factors $(z_1-y)^{2/\kappa}$ and $(z_2-y)^{2/\kappa}$ and by calculating
\beq
    \int_{z_1}^{z_2} dy
    (y-z_1)^{-\frc4\kappa}(z_2-y)^{-\frc4\kappa} =
    \frc{\G\lt(1-\frc4\kappa\rt)^2}{\G\lt(2-\frc8\kappa\rt)} (z_2-z_1)^{1-\frc8\kappa}
\eeq
which, multiplied by $(z_2-z_1)^{2/\kappa}$ coming from the factor $G_N^f$, gives
\beq
    \propto (z_2-z_1)^{1-\frc6\kappa} = (z_2-z_1)^{-2r_b}.
\eeq
This is the bosonic behavior. It is in fact a purely bosonic
behavior (corrections are positive integer powers of $z_2-z_1$), and the exact leading part of $G_{{\cal C}_i}$ is
given by (taking the normalisation as in (\ref{Gintg4real}))
\beqa
    G_{{\cal C}^{(1)}} &=& 2\;\sin\frc{4\pi}\kappa\; \frc{\G\lt(1-\frc{4}\kappa\rt)^2}{\G\lt(2-\frc{8}\kappa\rt)}\;
    G_{N-2}^f(\theta_3,\ldots,\theta_N)\; \lt(\sin\frc{\theta_1-\theta_2}2\rt)^{-2r_b}\times \n && \times
    \lt(1+O(\theta_1-\theta_2)\rt)~.\label{GC1a}
\eeqa
Here we wrote everything back in terms of the angular variables, and we wrote explicitly the dependence of
$G_{N-2}^f$ on these variables for clarity.
The behaviors as $z_2\to z_3$ and $z_N \to z_1$ are generically modified: they have a bosonic part and
a fermionic part. For the leading bosonic behavior as $z_2\to z_3$, for instance, one just replaces
the variables $z_2$ and $y$ by $z_3$, except in the factors $(z_2-y)^{-4/\kappa}$,
$(z_3-y)^{-4/\kappa}$ and in the integration limit $z_2$. Taking the integration limit $z_1$ to $\infty$
gives the leading bosonic behavior (again with the normalisation as in (\ref{Gintg4real}))
\beqa &&\label{GC1b}
    G_{{\cal C}^{(1)}} = 2\;\sin\frc{4\pi}\kappa\;
    \frc{\G\lt(1-\frc4\kappa\rt) \G\lt(-1+\frc8\kappa\rt)}{\G\lt(\frc4\kappa\rt)}\;\times\\ &&\qquad\times\;
    G_{N-2}^f(\theta_1,\theta_4,\ldots,\theta_{N})\; \lt(\sin\frc{\theta_2-\theta_3}2\rt)^{-2r_b}\
    \lt(1+O(\theta_2-\theta_3)\rt)+\n && \qquad + \ O\lt((\theta_2-\theta_3)^{-2r_f}\rt) \no
\eeqa
The fermionic part (which is subleading) has a more complicated expression that we will not write here, and
is generically non-zero. We expect that, from the viewpoint of the Calogero-Sutherland hamiltonian, the
behaviors (\ref{GC1a},\ref{GC1b}) (with the explicit constant for the fermionic behavior in (\ref{GC1b})) fixes the Hilbert space. Note that this is a different Hilbert
space than the usual fermionic or bosonic ones, hence the new ground state (\ref{Gintg4real}) does not violate the unicity of the known fermionic of bosonic ground states
of the Calogero-Sutherland hamiltonian.

It is worth noting, however,
that in the case $N=3$ and $\kappa=6$, the functions $G_{{\cal C}^{(i)}}$ all degenerate to constants
(that is, in this case all behaviors are purely bosonic); this is just the solution $G^b_3$.

\ssect{Many integration variables}

It is a simple matter to extend the method to integral formulas with many integration variables. Consider now
\beq\label{wM}
    w = G_N^f \prod_{1\leq j<k\leq M}
    \lt|\sin\frc{\zeta_j-\zeta_k}2\rt|^{-2\beta_{jk}} \hspace{-5mm}
    \prod_{1\leq j\leq N,\,1\leq k\leq M}\lt|\sin\frc{\theta_j-\zeta_k}2\rt|^{-2\alpha_k}.
\eeq
A calculation similar to that of the previous sub-section shows that this is an eigenfunction of the operators
\beq\label{WjM}
    {\cal W}_j = {\cal D}_j + \sum_{k=1}^M \lt(f_j^k\p_{\zeta_k} -
    (f_j^k)'\rt)
\eeq
where $f_j^k = f(\theta_j-\zeta_k)$, if and only if
\beq
    \beta_{jk} = -\kappa \alpha_j\alpha_k\quad\mbox{and}\quad  \lt[\  \alpha_j = -\frc12 \quad \mbox{or}\quad
      \alpha_j = \frc2\kappa \ \rt]
\eeq
for all $j\,,~k = 1,\ldots,M$ (and in general, we can have $\alpha_j\neq \alpha_k$ for $j\neq k$). Let us denote
by $Q$ the number of parameters $\alpha_j$ that are set to $-\frc12$, and by $R$ the number of parameters
$\alpha_j$ that are set to $\frc2\kappa$ (that is, $Q+R=M$). Then, the eigenvalue associated to $w$ is
\beq\label{dNQR}
    d_N^{(Q,R)} = \frc{1}{2\kappa}\lt(N-2R+\frc\kappa2 Q\rt)^2 - \frc{(\kappa-4)^2}{8\kappa}
\eeq (concerning the relation with our previous notation, we have
$d_N^{\lt(-\frc12\rt)} = d_N^{(1,0)},\,d_N^{\lt(\frc2\kappa\rt)} =
d_N^{(0,1)}$). Again, we can construct eigenfunctions of all
operators ${\cal D}_j$ with the eigenvalue above by considering
the analytic continuation of the function $w$ (on a branch of our
choice) and by constructing \beq\label{genintegformulazeta}
    G_{{\cal C}_1,\ldots,{\cal C}_M} = A \int_{{\cal C}_1} d\zeta_1 \cdots \int_{{\cal C}_M} d\zeta_M
    \; w
\eeq
where the contours ${\cal C}_1,\ldots,{\cal C}_M$ must be topologically non-trivial.
Introducing the variables $y_k = e^{i\zeta_k}$ and $z_j = e^{i\theta_j}$ will simplify the discussion
of the contours. We then have
\beqa
    w &=& G^f_N \;(2i)^{\upsilon} \;\prod_{1\leq k\leq M} y_k^{\alpha_k \lt(N-2R+\frc{\kappa}2Q + \kappa\alpha_k\rt)}
    \times \n && \times
    \prod_{1\leq j\leq N,\,1\leq k\leq M} \lt[z_j^{\alpha_k} (z_j-y_k)^{-2\alpha_k}\rt]\;
    \prod_{1\leq j< k\leq M} (y_j-y_k)^{2\kappa\alpha_j\alpha_k}\no
\eeqa
and
\beq\label{genintegformula}
    G_{{\cal C}_1,\ldots,{\cal C}_M} = A \int_{{\cal C}_1} \frc{dy_1}{y_1} \cdots \int_{{\cal C}_M} \frc{dy_M}{y_M}
    \; w
\eeq
with
\[
    \upsilon =
    -\frc{\kappa^2 Q (Q-1) + 4\kappa Q(N - 2 R) + 16 R(R-N-1)}{4 \kappa}~.
\]

\sssect{Case $R=0$: excited states above the fermionic ground state}

\begin{figure}[t]
\bc \includegraphics{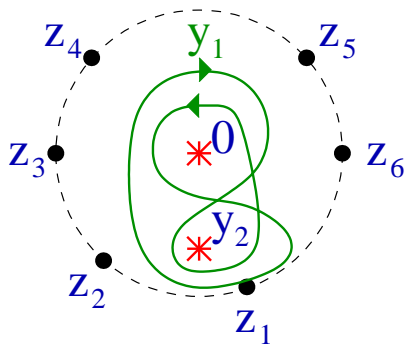} \ec
\caption{The $y_1$ part of the set of contours ${\cal C}^2_{{\rm origin}}$ (with $N=6$).}
\label{contour-double8}
\end{figure}

Taking $R=0$, there are no singularities at the points
$z_1,\ldots,z_N$, but there are at $y_j=y_k,\,j\neq k$ and at
$y_k=0$. The contours ${\cal C}_k$'s in $y_k$-planes can be made
non-trivial by ``surrounding themselves'' and surrounding the
origin, in a way that generalises the case $Q=1$ discussed in the
previous sub-section. For instance, one may first integrate over
$y_1$ surrounding the origin and the point $y_2$ in the ``double
8'' contour shown in Fig. \ref{contour-double8} (note that a figure-8 contour is not
closed in this case, since the origin and $y_2$ are algebraic
singularities with different powers). Then, one may integrate the
variable $y_2$ surrounding the origin and $y_3$, etc., until only
the variable $y_Q$ is left. The remaining integral is of the form
$\int dy_Q/y_Q\;y_Q^{-QN/2}[[y_Q]]$ (by power counting), which is
non-zero only if $QN$ is even; the contour can then be taken
surrounding the origin. This set of contours ${\cal
C}_1,\ldots,{\cal C}_Q$ will be denoted ${\cal C}^Q_{{\rm
origin}}$. One can check that explicit calculations give zero for
any odd $N$, hence the condition for having non-zero results is
that $N$ be even.

Following the discussion in the paragraph above (\ref{excfermionic}), a configuration
of $p_1,\ldots,p_{N-1}$ that reproduces the dimension $d_N^{(Q,0)}$ satisfies
\beqa
	\sum_{l=1}^{N-1} \sum_{l'=1}^{N-1} (N {\rm min}(l,l') - ll') p_l p_{l'} &=& \frc{N^2Q^2}4 \n
	\sum_{l=1}^{N-1} l(N-l) p_l = \frc{N^2Q}4~.\no
\eeqa
From this, it is simple to check that for all even $N$,
\beq\label{excfermionicQ}\ba{l}
    d^{(Q,0)}_N = d_N^{f;\,p_1,\ldots,p_{N-1}} \z \mbox{with} \quad \lt\{\ba{l}
    p_1=\cdots = p_{N/2-1} = p_{N/2+1} = \cdots = p_{N-1} = 0~,\\ p_{N/2}=Q \ea\rt. \z
    (N \ \mbox{even}). \ea
\eeq
For $N=2$ these states are not degenerate, and for $N=3$ there are no configurations of $p_1,p_2$ that would give $d^{(Q,0)}_3$. However, the situation is more complicated for higher $N$. For $N=4$, the states with $Q=3,6,9,12,\ldots$ are degenerate, hence there are other configurations of $p_1,\ldots,p_{N-1}$ giving
$d^{(Q,0)}_4$; for $N=5$ there are configurations for $Q=8,16,\ldots$. However, we expect that our solution (\ref{genintegformula}) with the contour ${\cal C}^Q_{\rm origin}$ gives the states described by (\ref{excfermionicQ}). A direct proof of this, for all $N$ and $Q$, is beyond the scope of this paper. The result of the integrals gives, for instance in the case $Q=2$ (up to a phase),
\beqa
    G_{{\cal C}^2_{{\rm origin}}} &=& A \;G^f_N \;(2i)^{\upsilon} \;
    \sum_{p=0}^N \frc{\G\lt(-\frc\kappa4-\frc{N}2+p\rt) \G\lt(\frc\kappa2+1\rt)}{\G\lt(\frc\kappa4-\frc{N}2+1+p\rt)}
    \times \n &&\times
    \sum_{u\subset\{1,\ldots,N\},\,v\subset\{1,\ldots,n\}\atop |u|=|v|=p} \prod_{j=1}^{N} \frc{z_{v_j}}{z_{u_j}}~.
\eeqa
One can see that the formula above is exactly zero for $N$ odd, as claimed.

We conjecture that with all integers $Q>0$,
the fermionic excited states of the
Calogero-Sutherland hamiltonian characterised by the sets $\{p_i\}$ as in
(\ref{excfermionicQ}) are all excited states above the fermionic ground state that
satisfy simultaneously all null-vector equations.

\sssect{Case $Q=0$: ground state with mixed boundary conditions}

Taking $Q=0$, there are singularities at the points
$z_1,\ldots,z_N$. In a fashion similar to what we did in the
previous sub-section, we can form pairs by integrating each of
the variables $y_j$ around two points $z_{i_j}$ and $z_{i_j+1}$ in
the figure-8 contour ${\cal C}^{(i_j)}$, in such a way that the
contours do not cross each other. Again, with $\kappa>4$, the
contours can be collapsed to lines joining adjacent angles. This
modifies the boundary conditions as in (\ref{formbc1}), for all
indices $i_1,\ldots,i_R$ involved: the condition is purely bosonic
when two angles of a same pair collide, it is mixed when the
angles are part of different pairs or one of them only is part of
a pair, and purely fermionic when both angles are not part of any
pairs.
There are other contours giving linearly independent functions (and other complicated boundary conditions),
but we will not analyse them here (we do not expect them, generically, to give rise to real positive solutions).
We believe that the contours described here may give rise to
functions $G_{{\cal C}_1,\ldots,{\cal C}_R}$ with a simple stochastic interpretation, that we develop
in the next section.

The expressions can be made obviously real and positive for $\kappa>4$, hence these are ground states
(here again, choosing $A = (1/2)^R e^{i\varphi}$ for some real $\varphi$ depending on the indices
$i_1,\ldots,i_R$):
\beqa\label{mixedgs} &&
    G_{{\cal C}^{(i_1)},\ldots,{\cal C}^{(i_R)}} = \lt(\sin\frc{4\pi}\kappa\rt)^R
    G^f_N~\int_{\theta_{i_1}}^{\theta_{i_1+1}} d\zeta_1\cdots
    \int_{\theta_{i_R}}^{\theta_{i_R+1}} d\zeta_R\;\times \\ && \qquad \times
    \prod_{1\leq j<k\leq M}
    \lt|\sin\frc{\zeta_j-\zeta_k}2\rt|^{\frc8\kappa}
    \prod_{1\leq j\leq N,\,1\leq k\leq M}\lt|\sin\frc{\theta_j-\zeta_k}2\rt|^{-\frc4\kappa} \n && \qquad
    (\kappa>4) \no
\eeqa
for $i_a\in\{1,\ldots,N\}$, $i_{a+1}-i_a \ge 2$ and if $i_1=1$ then $i_R<N$.
The expansions at colliding angles are of the same form as those of the previous sub-section
\beqa\label{bcbosgeneral}
    && G_{{\cal C}^{(1)},{\cal C}^{(i_2)},\ldots,{\cal C}^{(i_R)}} =
    2\;\sin\frc{4\pi}\kappa\; \frc{\G\lt(1-\frc{4}\kappa\rt)^2}{\G\lt(2-\frc{8}\kappa\rt)}\;\times\\ &&
    \qquad\qquad\qquad\times\;
    G_{{\cal C}^{(i_2)},\ldots,{\cal C}^{(i_R)}}(\theta_3,\ldots,\theta_N)\;
    \lt(\sin\frc{\theta_1-\theta_2}2\rt)^{-2r_b}\times\n && \qquad\qquad\qquad\times
    \lt(1+O(\theta_1-\theta_2)\rt) \no
\eeqa
and
\beqa && \label{bcgeneral}
    G_{{\cal C}^{(1)},{\cal C}^{(i_2)},\ldots,{\cal C}^{(i_R)}} = 2\;\sin\frc{4\pi}\kappa\;
    \frc{\G\lt(1-\frc4\kappa\rt) \G\lt(-1+\frc8\kappa\rt)}{\G\lt(\frc4\kappa\rt)}\;\times\\ &&
    \qquad\qquad\times\;
    \lt\{\ba{ll}G_{{\cal C}^{(i_2)},\ldots,{\cal C}^{(i_R)}}(\theta_1,\theta_4,\ldots,\theta_{N})
    & (i_2\neq 3) \\
    G_{{\cal C}^{(1)},{\cal C}^{(i_3)},\ldots,{\cal C}^{(i_R)}}(\theta_1,\theta_4,\ldots,\theta_{N})
    & (i_2= 3)\ea\rt\} \;\times\n &&
    \qquad\qquad\times\;
    \lt(\sin\frc{\theta_2-\theta_3}2\rt)^{-2r_b}\
    \lt(1+O(\theta_2-\theta_3)\rt)  +\n && \qquad\qquad+ O\lt((\theta_2-\theta_3)^{-2r_f}\rt)~. \no
\eeqa
Here we wrote explicitly the dependence on the angles where necessary for clarity.  Again, we expect that, from the viewpoint of the Calogero-Sutherland hamiltonian, the
behaviors (\ref{bcbosgeneral},\ref{bcgeneral}) (with the explicit constant for the fermionic behavior in (\ref{bcgeneral})) fixes the Hilbert space.

\sssect{Case $Q\neq0,\,R\neq0$: excited states with mixed boundary conditions}

Finally, it is a simple matter to combine the two types of
contours mentioned above in the general case. Functions of the
type \beq
    G_{{\cal C}^Q_{{\rm origin}},{\cal C}^{(i_1)},\ldots,{\cal C}^{(i_R)}}(\theta_1,\ldots,\theta_N)
\eeq
(again, for $i_a\in\{1,\ldots,N\}$, $i_{a+1}-i_a \ge 2$ and if $i_1=1$ then $i_R<N$) satisfy all null-vector
equations (\ref{nveq}), as well as the boundary conditions (\ref{formbc1}) for all indices
$i_1,\ldots,i_R$ involved. They describe certain excited states
above the mixed ground state corresponding to (\ref{mixedgs}), described in the previous paragraph,
generalising the excited states (\ref{excfermionicQ}).
Obviously, these are not the only contours that can be taken, but we believe that these contours give
rise to functions $G_{{\cal C}^Q_{{\rm origin}},{\cal C}^{(i_1)},\ldots,{\cal C}^{(i_R)}}$ that describe
all possible (zero-momentum) excited states above the mixed ground states (\ref{mixedgs}) that can be obtained
from our general integrale formula (\ref{genintegformula}) (but note that the construction
that we explain in the next sub-section gives yet other zero-momentum excited states).
Also, they are not all possible excited states above
this mixed ground state, but we conjecture that they are all (zero-momentum)
excited states that satisfy simultaneously all null-vector equations.

\ssect{Solutions with non-zero total momentum or spin}
\label{sectspin}

A solution to the Calogero-Sutherland eigenvalue equation with non-zero total momentum
is simply obtained by multiplying a zero-momentum solution
by the exponential $e^{is\sum_j \theta_j}$.
The energy then gets added by the term $\frc{Ns^2 }2$,
and the total momentum is just $Ns$. Generally, multiplying by this exponential factor a solution of
energy $E$ and momentum $P$ gives a new solution of energy $E+\frc{Ns^2}2 + sP$ and momentum $P+Ns$.
Hence this corresponds to adding a momentum $s$ to that of each particle (making the quantum-mechanical average
of the momentum of each particle exactly what it was before plus the value $s$), and the multiplication by this
phase factor is just the Galilean transformation of the initial eigenfunction.
Note that in order for the eigenfunction to be still defined on the circle, the total momentum must be an integer,
hence we must have $Ns \in \Z$ (otherwise, one may in fact interpret the particles as anyons confined on a circle).

The total momentum operator $-i\sum_j \p_j$ is, on correlation functions, the operator for the spin
of the bulk field. This would suggest that we would be able to construct in this way correlation functions
corresponding to bulk fields with non-zero spin.
However, such Galilean-transformed eigenfunctions do not generically give rise to solutions to all null-vector
equations, since the operators ${\cal D}_j$ transform non-trivially under Galilean transformation.
There is a way, though, to obtain solutions to the null-vector equations that carry a non-zero spin,
corresponding to non-zero total momentum for the eigenfunctions. Consider the transformation property
\beq
    e^{-is\sum_j \theta_j} {\cal D}_j e^{is\sum_j \theta_j} =
    {\cal D}_j - is\kappa \p_j + is F_j + \frc{\kappa s^2}2~.
\eeq
With
\[
    \Gamma = \sum_j \theta_j + \gamma \zeta
\]
for some number $\gamma$, we then have
\beq
    e^{-is\G} \cdot {\cal W}_j \cdot e^{is\G} = {\cal W}_j - is\kappa \p_j + isF_j + \frc{\kappa s^2}2 + is\gamma f_j
\eeq
where ${\cal W}_j$ is defined in (\ref{Wj1}). Note that
\[
    (\kappa \p_j - F_j - \gamma f_j) w = -(\kappa \alpha+\gamma) f_j w
\]
where $w$ is defined in (\ref{w}). Hence, choosing
\beq
    \gamma = -\alpha\kappa = \lt\{\ba{ll}  \frc\kappa2 & \lt(
    \alpha=-\frc12\rt) \\  -2 & \lt(\alpha=\frc2\kappa\rt) \ea\rt.
\eeq
gives
\beq
    {\cal W}_j \lt(e^{is\G}w\rt) = \lt(\lambda + \frc{\kappa s^2}2\rt) e^{is\G} w
\eeq
where $\lambda$ is the eigenvalue (\ref{EigenWj1e}) or (\ref{EigenWj1}).

Hence, a solution with non-zero
spin is obtained by replacing $w$ by $e^{is \G}w$ in the integral expression (\ref{integexpr1}) above.
The field dimension gets added by the term $\frc{\kappa s^2}2$, and the energy, by the term $\frc{Ns^2}2$.
The same structure works for many integration variables: one needs to replace $w$
by $e^{is(\sum_j \theta_j - \kappa\sum_k \alpha_k\zeta_k)}w$ in
(\ref{genintegformulazeta}), and the field dimension and energy get added by the same terms.
Let us denote generically the resulting correlation function by $G^{(Q,R)}_s$, employing
the notation $Q$ and $R$ as in the paragraph above (\ref{dNQR}). Then, the spin of the bulk field,
equivalently the total momentum of the eigenfunction, is given by $\lt(N + \frc{\kappa Q}2 - 2R\rt)s$:
\beq
    -i\sum_{j=1}^{N} \p_j G^{(Q,R)}_{s} = \lt(N + \frc{\kappa Q}2 - 2R\rt)s\; G^{(Q,R)}_{s}~.
\eeq
Note that it is {\em not} just $Ns$.

When $Q=0$, it is like giving momentum $s$ to each of the $N-2R$ fermions
(the particles that are not paired by bosonic boundary conditions), and giving no average momentum to the bosons
(the particles that are paired). We believe that it is indeed what happens
if the contours are chosen as in the discussion in the previous sub-sections.
These contours are still valid, since
the singular points surrounded by the integration contours of the $R$ variables with
$\alpha=\frc2\kappa$ are not affected by the extra factors coming from $e^{is\G}$. However, in the case $Q=0$,
there are still other non-trivial contours for discrete ranges of $s$ (for instance, with $R=1$,
a necessary condition is: whether $2N/\kappa -2s \in \Z$ and $s\leq N/\kappa$,
or $-2N/\kappa -2s \in \Z$ and $s\geq -N/\kappa$): the
contours ${\cal C}_{\rm origin}$ that circle the origin, or similar contours that circle the point at infinity.
These contours do not affect the boundary conditions, and if $R-R'$ variables are taken with
such contours, we have new non-zero-momentum solutions with $2R'<2R$ bosons and $N-2R'>N-2R$ fermions.

When $Q\neq0$, the situation becomes even more complicated. The $Q$ integration
variables associated with $\alpha=-\frc12$ now live on Riemann surfaces with more complicated singularity
structures at the origin and at infinity. The conditions for having non-trivial integration contours are
generically affected. For instance, in the case $Q=1$ and $R=0$, a necessary condition is
$\kappa s/2\equiv q \in \Z+N/2$ and
$-N/2\leq q \leq N/2$. The general case $Q\neq0,R\neq0$ should comprise a myriad of contours, including
those we described in the previous sub-sections as well as those we described here, along with conditions
on the spins.

Let us note here that in order for the eigenfunction to be well-defined on the circle we need
\beq\label{quant}
    \lt(N + \frc{\kappa Q}2 - 2R\rt)s \in \Z~.
\eeq
It is important to realise that this condition may not be in agreement with the conditions on $s$ that arose
above for having certain non-trivial contours. However, it is always in agreement with the contours
of the previous sub-sections in the cases where $Q=0$. That is, the eigenfunction with only the $N-2R$ fermions
being given an average momentum $s$ is a valid one.

The condition (\ref{quant}) is not really necessary from the viewpoint of correlation functions:
it is conceivable that a correlation function acquires a phase when the
positions of the null fields are all brought around the circle (this would correspond to the bulk field
being ``semi-local'' with respect to the boundary null-fields).

From the viewpoint of eigenfunctions of the Calogero-Sutherland
hamiltonian, we can now apply a Galilean transformation to bring
the momentum back to zero, and we obtain {\em different}
zero-momentum solutions with a {\em different} energy from those
corresponding to $G^{(Q,R)}_{0}$. These should not correspond to
ground states, because they have no reason to be real and
positive. The energy is given by (we denote by $E_N^{(Q,R)}$ the
energy corresponding to the dimension $d_N^{(Q,R)}$ defined in
(\ref{dNQR})) \beq
    E_N^{(Q,R)}  + \frc{Ns^2}2\lt[1 - \lt(1+ \frc{\kappa Q}{2N}-\frc{2R}N \rt)^2\rt]
\eeq
We do not fully understand yet the meaning of these new zero-momentum solutions. Certainly,
for $R=0$ these are yet other excited states above the fermionic ground states; hence we have here integral
representations for these other excited states (and these should agree, of course, with the known
eigenfunctions). We have not fully identified them, because we have not fully
determined the conditions on $s$ for all $Q>0$.

With $R\neq0$ and $Q\neq0$, we obtain new excited states above the ground states with mixed boundary conditions
by taking the contours of the $R$ integration variables associated to $\alpha=\frc2\kappa$ as in the discussion
in the previous sub-sections, and with $s\neq0$ restricted by the conditions coming from the
$Q$ integration variables associated to $\alpha=-\frc12$.
But, we can also take some of the $R$ variables to have contours surrounding
the origin or infinity, as described above, as long as the resulting conditions on the spin are in agreement
with those coming from the $Q$ variables associated to $\alpha=-\frc12$ (and if $Q=0$, there is no agreement
conditions). We obtain new zero-momentum
excited-states eigenfunctions with $2R'<2R$ bosons and $N-2R'>N-2R$ fermions.

The case $Q=0$ seems at this point slightly problematic:
by taking the contours as in the previous sub-sections,
we obtain new excited states above the $2R$-boson, $N-2R$-fermion ground states, with energies
\beq
    E_N^{(0,R)}  + \frc{Ns^2}2\lt[1 - \lt(1-\frc{2R}N \rt)^2\rt]
\eeq
that form a continuum. Indeed, here $s$ does not seem to be restricted by any condition for having non-trivial
contours, and since the wave function has zero momentum, there are no conditions coming from imposing that it
be defined on the circle. We do not know how to interpret this continuum of zero-momentum solutions,
if really it occurs; a more involved analysis
of the explicit integral formulas would certainly be useful for this purpose.

\ssect{Interpretation via Coulomb gas formalism of CFT}

The goal of this sub-section is to clarify our construction in relation to the Coulomb gas formalism of CFT.

In the Coulomb gas formalism, one first constructs (boundary) vertex operators $V_p(\theta)$
(they are operators that act on the Hilert space of radial quantization of CFT)
with dimensions $p^2-2pq$ for some fixed $q$, and with ``charge'' $p$. The charge of a vertex
operator is just the associated eigenvalue of a charge operator $Q$ (that is,
$[Q,V_p] = pV_p$), that is supposed to exist on the Hilbert space. This means
that the product $V_{p_1} V_{p_2}$ has charge $p_1+p_2$, and taking into consideration the dimension, we have
the OPE's
\[
    V_{p_1}(\theta_1) V_{p_2}(\theta_2) \sim (\theta_1-\theta_2)^{2p_1p_2} V_{p_1+p_2}(\theta_2) ~.
\]
The characteristic properties of a these vertex operator
is that correlation functions of products of such objects
are non-zero only when the total charge is equal to $2q$, and that
they evaluate, in our context, to the product of all pairings of the vertex
operators involved, a pairing being just equal to
$[2\sin((\theta_1-\theta_2)/2)]^p$ where $p$ is the power of $\theta_1-\theta_2$ that appears in the leading OPE.

One then constructs certain special level-2 null fields $\phi$ (and higher-level null fields as well) by choosing
$q=\frc{\kappa-4}{4\sqrt{\kappa}}$ and identifying $\phi= V_p$ with $p=\frc1{\sqrt{\kappa}}$.
This indeed reproduces the correct OPE's of such null fields, but without the identity component; hence,
these are very special level-2 null-fields.

In our case, we take these boundary fields and put at the center of the disk
the product of bulk holomorphic and anti-holomorphic vertex operators
$V_{P} \b{V}_{P}$, with charge $P = -Np/2+q$ if there are $N$ boundary
null fields. By mapping the boundary theory to a holomorphic theory on the full plane (where
$\b{V}_P$ becomes $V_P$ at infinity), we are left with correlation functions of holomorphic vertex operators.
The total charge requirement is satisfied, hence correlation functions are non-zero.
This indeed reproduces the fermionic correlation function $G_N^f$ (with an appropriate normalisation),
and in particular, one can check that
the dimension of the product of bulk vertex operators is $d_N^f$.

The Coulomb gas construction then goes on to construct more complicated null-fields by inserting a
vertex operator of dimension 1 and by integrating its position over closed contours. This insertion
scales as a dimension 0 non-local object, and its effect is to change the fields that are involved.
More precisely, the null fields become different null fields (with different OPE's that contain the identity
field), and the bulk field is modified. There are two possible dimension-1 vertex operators:
$V_{r_\pm}$ with $r_\pm  = q \pm \sqrt{q^2+1}$. The correlation function (\ref{w}) is exactly the correlation
function with one such insertion, and the function (\ref{wM}) is the correlation function with
$M$ insertions. Our derivation shows how these insertion modifies the fields for various choices of the
contours.

\sect{Completeness in the case $N=3$ spinless}
\label{N3}

In this section, we consider in general the problem of determining
if an eigenfunction of the Calogero-Sutherland hamiltonian can
give a boundary CFT correlation function $G$ in the case of 3
particles. We show that mixed boundary conditions with $N=3$
impose the dimension of the field to be $d_1^f$ and, in the cases
where $\kappa\neq 6$ and $\kappa \neq 8/n$ with $n=2,3,\ldots$, we
argue that the solution with any kind of mixed boundary condition
has a 3-dimensional basis composed by the solutions with purely
bosonic behaviors at some pair of colliding angles that we
described in the previous section.

We will start with considerations for general particle number $N$.
For definiteness, consider the ordering of the angles to be
$\theta_1>\theta_2>\cdots>\theta_N$ and consider the
behavior as $\theta_1\to\theta_2$: it is a linear combination
of the power laws $(\theta_1-\theta_2)^{-2r_f}$ and $(\theta_1-\theta_2)^{-2r_b}$.

As we said in the Introduction,
the constraints that come out of the system of equation (\ref{nveq}) are essentially part of the
general theory of null-vectors in CFT. In particular, the fermionic behavior (\ref{fbc}) corresponds
to a fusion into a level-3 degenerate boundary field ($\phi_{1,3}$ in the Kac classification),
and the bosonic behavior
(\ref{bbc}) to a fusion into the identity ($1$) operator.

\ssect{Conditions from null-vector equations}

It will be convenient to consider the separation $\Delta$ between $d_\Phi$ and the 1-leg exponent
$d_1^f$ (\ref{Nlegexp}), that is, the equations ${\cal D}_j G = (d_1^f + \Delta) G$.
We first look at arbitrary $N$.
A generic solution to the Calogero-Sutherland system has, around $\theta_1=\theta_2$, a basis of the form
\beq
    G = \theta_{1,2}^{-2r} (A+B\theta_{1,2} +
    C\theta_{1,2}^2 + D\theta_{1,2}^3+\ldots)
\eeq
where $r=r_f$ or $r=r_b$, where we use
\beq
    \theta_{j,k} = \theta_j-\theta_k
\eeq
and where $A\neq0,\,B,\ldots$ are functions of $\theta_2,\ldots,\theta_N$. (That is, in general
$G$ can be a linear combination of one expansion with $r=r_f$ and one with $r=r_b$.)
Note that for $\kappa = 8/n$ with $n=1,2,3,\ldots$, we have ``resonances'': $-2r_f = -2r_b + n$;
however, we will not look at the resulting logarithmic behaviors.

As shown in \ref{app2}, the null-vector equations (\ref{nveq}) lead to the following constraints,
which can as well be seen as coming from null-vector CFT considerations:
\beq\label{cond1}\ba{l}
    (r = r_f \quad\mbox{and}\quad \p_2 A = 2B) \z  \mbox{or} \quad ( r=r_b \quad\mbox{and}\quad
    \p_2 A = 0 \quad\mbox{and}\quad (B=0 \quad\mbox{or}\quad \kappa=4))~,\ea
\eeq
\beq\label{1.2}\ba{l}
    \sum_{k\neq1,2} (f_{2k}\p_k -hf_{2k}')A - \frc16
    (2r-h)A+\z+ 2\p_2 B - \frc{(2r\kappa-\kappa-6)(r \kappa-\kappa+1)}\kappa\, C = \Delta A\ea
\eeq
and
\beqa\label{3.2}
    r=r_b &\Rightarrow& -\frc{\kappa}2 \p_j^2A + \sum_{k\neq1,2,j} (f_{jk}\p_{k} - hf_{jk}')A
      = \Delta A \\
\label{3.3}
    r=r_f &\Rightarrow& -\frc{\kappa}2 \p_j^2A + \sum_{k\neq1,2,j} (f_{jk}\p_{k} - hf_{jk}')A
     +\n && + f_{j2}\p_{2} A - 2h_{3,1}f_{j2}'A = \Delta A~.
\eeqa
Eq. (\ref{3.2}) is the equation ${\cal D}_j^{(N-2)} A = (d_1^f + \Delta) A$ with
the differential operator ${\cal D}_j^{(N-2)}$
being like ${\cal D}_j$ but for the $N-2$ angles $\theta_3,\ldots,\theta_N$,
instead of the $N$ angles. Also, in (\ref{3.3}),
$h_{1,3} = \frc{8-\kappa}{2\kappa}$ is the dimension of a level-3 degenerate field.
Equations (\ref{3.2}) and (\ref{3.3}) indicate that the function $A$ describes, in the case $r=r_b$,
a correlation function with $N-2$ level-2 degenerate boundary fields, and, in the case $r=r_f$,
a correlation function with one level-3 degenerate
boundary field (at $\theta_2$) and $N-2$ level-2 degenerate boundary fields.

According to (\ref{cond1}), when the boundary fields fuse to the identity, the second term
of the expansion, with coefficient $B$, is absent, except possibly when $\kappa=4$. The case $\kappa=4$
corresponds to the theory with $c=1$, which is the free massless boson, where there is a natural operator
of dimension 1 which indeed occurs as a symmetry descendant of the identity operator.

Along with conditions (\ref{cond1}), equation (\ref{1.2})
fixes $C$ in terms of the function $A$ (and the number $\Delta$)
in the fermionic and bosonic cases with $\kappa\neq4$. For $\kappa=4$, the function $C$ also depends upon
$B$, which is not necessarily zero.

\ssect{Case $N=3$}

We now analyse in more detail the case $N=3$.
For spinless $\Phi$, the function $A$ depends only upon $\theta_2-\theta_3$.

Let us first analyse the bosonic case. Then, equation (\ref{3.2}) implies that $\Delta=0$ since $A$ is constant.
Further, equation (\ref{2.1}) fixes $C$ uniquely (up to normalisation):
\[
    C =
    \frc{h}{8-3\kappa}\lt(f_{23}'+\frc16\rt)
    A~.
\]
It is simple to see that all coefficients $D,\ldots$ are then also fixed uniquely,
if the solution exists, as long as $\kappa \neq 8/n$ for $n=2,3,4,\ldots$. It is worth noting that
since a solution with one purely bosonic behavior as some pair of colliding angles is unique, then
a solution with two purely bosonic conditions must have all bosonic conditions by cyclic permuttations.

When $\kappa=4$, still in the bosonic case, a further analysis shows that in fact
we must have $B=0$, so that $C$ and all other coefficients are also fixed uniquely, if the solution exists.
Then, there cannot be non-logarithmic solutions with purely fermionic behavior (and $\Delta=0$), since the fermionic
exponent occurs in the bosonic solution (a resonance). Hence, any non-logarithmic solution must be purely bosonic
everywhere,
but we showed that such solutions to the Calogero-Sutherland system do not satisfy (\ref{nveq}) at $\kappa=4$.
That is, a bosonic solution will also have to involve logarithms.

When $\kappa=8/3$, equation (\ref{2.1}) becomes inconsistent, as it requires $A=0$: a
bosonic behavior for this value of $\kappa$ will have to involve some logarithms as well.

Similarly, when $\kappa=8/n$ for $n=4,5,6,\ldots$, difficulties appear when trying to determine the coefficients
$D,\ldots$, and logarithms will be necessary.

Finally, it is worth noting that for $\kappa=6$, since
the solution is unique, it is given by the constant solution $G=const.$.

Hence, we have showed that for $\kappa \neq 8/n$ for $n=2,3,\ldots$,
if a general solution has some contribution to a bosonic behavior at any pair of colliding angles,
say at $\theta_1=\theta_2$, it must correspond to an operator
of dimension $d_1^f$, and that the purely bosonic contribution at $\theta_1=\theta_2$ is unique up
to normalisation. In the previous section, we constructed explicitly the unique solutions that are purely bosonic
as some pair of colliding angles, and we saw that, except for $\kappa=6$, they have fermionic contributions
at other pairs of colliding angles.
Now, the part of a general solution that is purely fermionic at $\theta_1=\theta_2$ is not uniquely fixed.
This part is fixed once the value of the function $A$ (involved in its expansion) is fixed.
But the function $A$ is ruled by (\ref{3.3}),
which determines the possible behaviors at colliding angles $\theta_2-\theta_3=0,2\pi$ in accordance to the
fusion rules $\phi_{1,3}\times \phi_{1,2} \mapsto \phi_{1,4}$ and $\phi_{1,3}\times \phi_{1,2} \mapsto \phi_{1,2}$.
Moreover, one can see that a choice of the ratios $V$ between the amplitudes of these two behaviors
as $\theta_2\to\theta_3^+$, for instance, along with the eigenvalue $\Delta$ completely fix the solution
up to a normalisation. Since we have $\Delta=0$, we are left with a one-dimensional
subspace of solutions for $A$ (up to normalisation). We already know of such a one-dimensional subspace:
it comes from taking linear combinations of
the particular solutions $G$ (constructed in the previous section) with purely
bosonic behavior at $\theta_2=\theta_3$, those with purely bosonic behavior at $\theta_3=\theta_1$, and
those with purely bosonic behavior at $\theta_1=\theta_2$, with the constraint that the behavior at
$\theta_1=\theta_2$ of the resulting linear combination is purely fermionic.
Hence, this constitutes all possible solutions with $\Delta=0$ that are purely fermionic
at $\theta_1=\theta_2$. In other words, any general solution $G$
that has some part of a bosonic behavior at some
colliding angles should be a linear combination of the three unique solutions that have purely bosonic
behavior at the three different pairs of colliding angles. This argument breaks down at the value $\kappa=6$,
since then only $G=const.$ can have pure bosonic behavior at some colliding angles.

\sect{Discussion}
\label{sectDiscuss}

As we mentioned, our results are based solely on level-2 null
vector equations of boundary CFT. Here, we attempt an
interpretation of our results as measures in the continuum $O(n)$
loop model at criticality \cite{Nienhuis82} (which we recall
below), mainly based on the values of the exponents that we found.

In order to calculate
prescribed measures from the system of differential equations,
one needs to specify the boundary conditions: the various proportion of bosonic and fermionic behaviors
at different pairs of colliding angle. Two problems arise.

The general problem of determining what boundary conditions completely fix
the solutions is quite involved; in the case $N=2$ it is the (solved!) problem of boundary conditions
for second-order differential equations, and in the case $N=3$ we solved it above (although not to
mathematical rigor). In the general case, it is related to finding OPE's of null fields that form a consistent
operator algebra. It is believed \cite{BauerBK05} that if one specifies all pairs, say $P$, where a bosonic behavior
occurs, along with some normalisation condition, then the solution to all null-vector equations is
fixed, and in particular the fermionic components at the pairs $P$ is fixed.
But in our case, the boundary conditions are specified
quite differently (in particular, we specify both bosonic and fermionic components at many pairs).

More importantly, the problem of relating a set of boundary conditions (or an operator algebra)
to prescriptions on measures is still far from being solved.
Very natural arguments were given in \cite{BauerBK05} for the case where no bulk field is present,
from SLE ideas. We will use these arguments below in a simplified version and
in a different language (without using SLE ideas) along with some of our solutions in order to derive conjectures for certain measures in the $O(n)$ model.

\ssect{Overview of the $O(n)$ loop model}

A measure in the lattice $O(n)$ model has the form
\beq\label{meas}
    \sum_{{\cal G}} {\rm x}^{\ell} n^{\omega}
\eeq
where ${\cal G}$ denotes all configurations of self and
mutually avoiding loops and, possibly, curves with prescribed
end-points on the honeycomb lattice, $\ell$ is the total length of
a configuration, $\omega$ is the total number of loops of a
configuration, and ${\rm x}=1/\sqrt{2+\sqrt{2-n}}$
\cite{Nienhuis82} is the value at which the system is critical (on
the honeycomb lattice), for real numbers $-2<n<2$. We will
imagine restricting all loops and curves to lie inside a disk. The
continuum limit is obtained by taking the lattice mesh size
infinitely small (equivalently, by taking the disk infinitely
large in units of lattice spacing). When the continuum limit of
this model is taken, it is expected, if it exists, to be described
in some way by a conformal field theory with central charge $c$
given in (\ref{paramch}) where $\kappa$ is related to $n$ via \beq
    n = 2\cos\, \pi\lt(1-\frc4{\kappa}\rt)~.
\eeq
In the continuum limit, the resulting curves are expected to be described by SLE$_\kappa$ curves.
For the sake of keeping the discussion concise, in the following we will not think in terms of SLE curves
(or in terms of growing such curves from some arbitrary points), but rather we will simply draw our intuition from
the idea of the continuum limit of the $O(n)$ model.

Taking the continuum limit of a certain measure of the $O(n)$ model requires an appropriate re-normalisation
of this measure. For instance, the measure (\ref{meas}) for configurations with only
one curve (apart from the loops) that starts and ends on fixed points on the boundary of the disk
becomes infinite as the mesh size is made smaller.
The series of numbers obtained as the mesh size is made smaller are quite meaningless. But if we
take the ratio of that measure with another measure where the curve starts and ends on {\em different}
fixed points, then we expect the limit of zero mesh size to be finite. This ratio is expected to be equal, in the
limit, to a ratio of correlation functions in CFT, where level-2 null fields are inserted
on the boundary of the disk at the positions where the end-points of the curves lie.
We will normalise measures for single curves starting
and ending at fixed points by always taking the ratio with, say,
the measure where the fixed points are exactly opposite each other on the boundary of the disk. The result is what
we will refer to as a measure on such configurations in the continuum, and is what corresponds to correlation
functions of null fields
up to a positive (non-zero) normalisation. For more curves and other prescriptions on their shapes,
we will keep the same principle: a measure will be a limit that only encodes the dependence on the starting
and ending points of the curves, obtained by taking the ratio with such a measure where
end-points are at arbitrarily chosen fixed positions. The limit is that of mesh size going to zero, and then
of other parameters going to zero if necessary for the definition of the bulk field or of new boundary fields.
The results of such normalised limits correspond to correlation functions.

It is worth noting that in the continuum $O(n)$ loop model, one can define fields
$\Orr_{n'}(x)$ by the fact that, in the underlying lattice model, loops
around the point $x$ are counted with the value $n'$ replacing
the value $n$ in the partition function. The dimensions of these fields was calculated in \cite{Cardy00},
and are given by
\beq\label{dimnprime}
    d_{n',n} = \frc{(\kappa'-\kappa)(\kappa\kappa' - 2\kappa-2\kappa')}{\kappa(\kappa')^2}
\eeq
where $n' = 2\cos\, \pi\lt(1-\frc4{\kappa'}\rt)$ (this will be used in the discussion in \ref{N2}).

\ssect{Fusion and measures}

As two angles collide, two power law behaviors for the measure are possible (here we disregard possible
logarithmic behaviors and resonances), and generically occur in linear combinations. The coefficient of the leading
term of each power law is itself another measure, for different curve configurations.
These new measures correspond to new correlation functions, where the two level-2 null fields have
been replaced by a single field, as occuring in their fusion:
$\phi \cdot \phi = {\bf 1} + \phi_{1,3}$. The field ${\bf 1}$ is the identity field and is associated to
the bosonic behavior, and $\phi_{1,3}$ is a level-3 null field and is associated to the fermionic behavior.
It is important to realise that only the fact that the field $\phi$ gives rise to the level-2 null vector equations
(\ref{nveq})) implies that ${\bf 1}$ is the identity and that $\phi_{1,3}$ is a level-3 null field
whose properties are essentially expressed in
(\ref{3.1}, \ref{3.2}, \ref{3.3})). What curve configurations are associated to these correlation functions
obtained by fusion?

\begin{figure}[t]
\bc \includegraphics{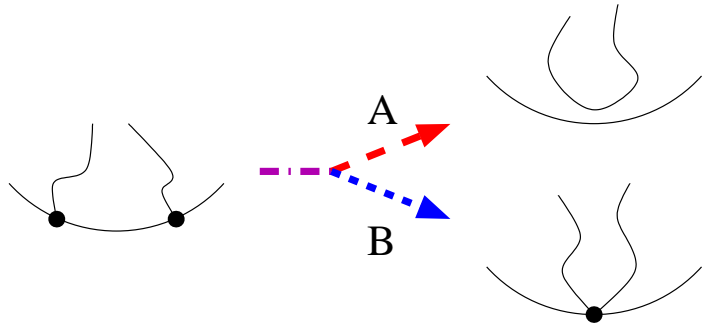} \ec
\caption{Interpretation of bosonic (A) and fermionic (B) behaviors in a generic boundary condition
involving a linear combination of these behaviors.}
\label{figbc}
\end{figure}

We give here only heuristic arguments. First, the measure
resulting from the fusion to the identity is that on
configurations where the curves touching the boundary at the
colliding angles are ``disconnected'' from the boundary and
joined, near to the boundary, into one curve (Fig. \ref{figbc} A).
Of course, such an operation is quite unprecise, but we will only
discuss qualitative features. Note that in order for this fusion
to occur, it must be, in some sense, that the resulting curve does
not affect the measure that corresponds to the correlation
function obtained by removing the two colliding null fields,
except for a possible normalisation (since the operator resulting
from the fusion is the identity).

Second, the measure resulting from the fusion to the level-3 null field
is that on configurations with the additional prescription that two curves start at the fused point
(Fig. \ref{figbc} B).

In \ref{N2}, we use these general ideas to explain various known exponents for the case $N=2$.

\ssect{The fermionic ground states and the $N$-leg exponents}

\begin{figure}[t]
\bc \includegraphics{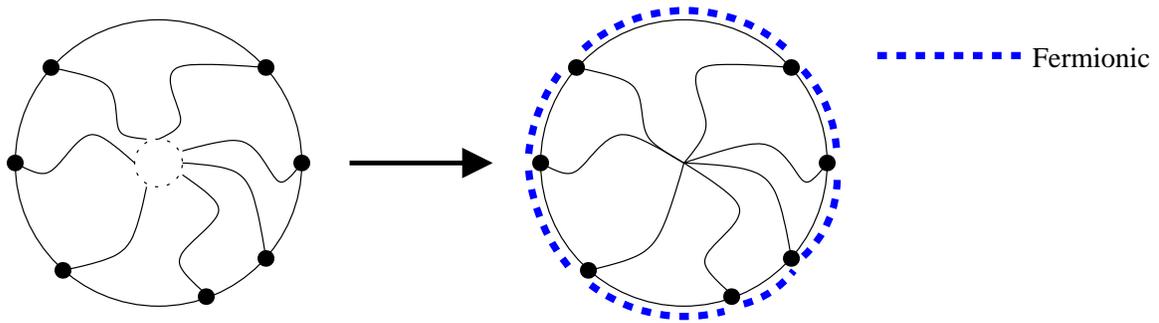} \ec
\caption{A typical configuration for the measure giving the $N$-leg exponent with $N=7$.
The inner circle has radius $\varep$, and the leading part as it is sent to 0 is the correlation function
$G_7^f$ times the power law $\varep^{d_7^f}$. The correlation function has all fermionic boundary conditions.}
\label{figNleg}
\end{figure}

The correlation function $G^f_N$ (\ref{Nlegcorrfct}) can be interpreted using the fact that the associated
bulk field dimension $d^f_N$ (\ref{Nlegexp}) is the $N$-leg exponent \cite{SaleurD87}.
That is, consider the measure on $N$ curves that have end-points at the angles
$\theta_1,\ldots,\theta_N$ on the boundary of the disk and at some angles (such that curves
don't cross each other) at a radius $\varep$ from the origin, as depicted in Fig. \ref{figNleg}.
As $\varep$ is sent to zero, this measure vanishes as $G_N^f\,\varep^{d^f_N}$, up to a
normalisation.

\begin{figure}[t]
\bc \includegraphics{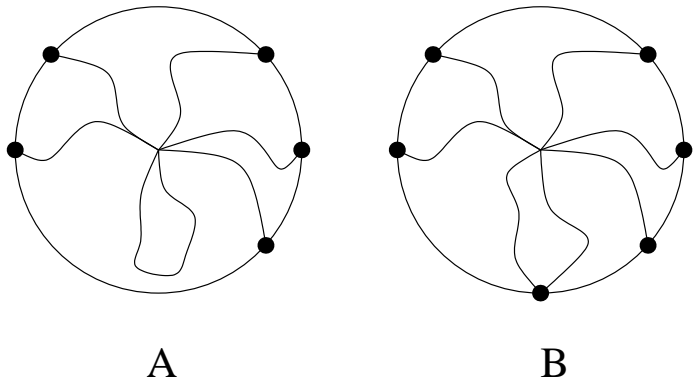} \ec
\caption{A typical configuration for the measure that would result as the coefficient of A) a bosonic behavior and B)
a fermionic behavior in the measure
giving the $N$-leg exponent with $N=7$, as the two downmost end-points are approached in Fig. \ref{figNleg}.
The bosonic behavior is forbidden, since the extra loop connected to the origin
affects the measure, but the fermionic
behavior is allowed since the two curves connecting a common point to the origin
extend macroscopically into the disk.}
\label{figNlegforbid}
\end{figure}

The fact that this situation corresponds to uniform boundary conditions of fermionic
type is easy to understand. Indeed, the bosonic behavior is divergent when $\kappa<6$,
but there is no reason, as two points collide, for the measure to grow. On the contrary, it should decrease
since there are less and less configurations as the curves emanating from the colliding points are more and more
constrained by each other. Hence there is no bosonic behavior. Another way of understanding, valid
for any $\kappa$, is as follows.
As two angles collide, the curve resulting from a bosonic behavior would be one that starts and
end on the center and that go all the way to a region not far from the boundary. But a curve that starts
and ends at the center has overwhelming probability of staying near the center; imposing that it goes to
a region near the boundary changes the measure, hence the fusion cannot give the identity operator; this is inconsistent, so there cannot be
bosonic behavior (see Fig. \ref{figNlegforbid}).

\ssect{The solutions with mixed boundary conditions and the $N'$-leg exponents}

The correlation functions $G_{{\cal C}^{(i_1)},\ldots,{\cal C}^{(i_R)}}$ (\ref{mixedgs})
should be interpreted similarly using the fact that the associated
bulk field dimension $d^f_{N-2R}$ (\ref{Nlegexp}) is the $(N-2R)$-leg exponent. One could then
expect that it is the measure on configurations like that of Fig. \ref{figintroconf} (at least
for $2R<N$; we will come back to the 0-leg exponent). Arguments like those
of the previous sub-section indeed suggest, for $\kappa<6$, that as two paired angles (two bosons, paired
by a curve joining them in the continuum $O(n)$ model) collide, the measure should grow very much,
since the curve can be made smaller and smaller. Hence, there should
be a bosonic behavior. However, it also suggest that the boundary condition
as a paired angle (a boson) collides with an unpaired angle (a fermion) is the purely fermionic one:
indeed, there is no reason for the measure to grow there, it should just decrease, as the curves
are constrained by each other. This suggests that our solutions (\ref{mixedgs}) are not the correct ones,
and that we have to take linear combinations of these solutions to obtain the desired behaviors (if possible).

When there is just one pair of bosons and $N$ particles, it is indeed possible to take linear combinations of our
$N$ independent solutions (where $N$ different pairs are taken) to obtain the suggested behavior: bosonic at
a single pair, say at the collision $\theta_1 \to\theta_2$, with some fermionic component, and purely fermionic everywhere else.
For a bosonic behavior at the collision $\theta_1\to \theta_2$, the linear combination $M_N^{(1)}$ is obtained from the inverse of a $N$ by $N$ matrix through
\[
    M_N^{(1)} \propto (1,0,0,\ldots,0) \mato{ccccccc} A & B & 0 & 0 & \cdots & 0 & B \\ B & A & B & 0 & \cdots & 0 & 0 \\
	0 & B & A & B & \cdots & 0 & 0 \\ \cdots & \cdots & \cdots & \cdots & \cdots & \cdots & \cdots \\
	B & 0 & 0 & 0 & \cdots & B & A \matf^{-1} \mato{c} G_{{\cal C}^{(1)}} \\ G_{{\cal C}^{(2)}} \\ \cdots \\ G_{{\cal C}^{(N)}} \matf
\]
where
\[
    A = 2\;\sin\frc{4\pi}\kappa\; \frc{\G\lt(1-\frc{4}\kappa\rt)^2}{\G\lt(2-\frc{8}\kappa\rt)}~,\quad
    B = 2\;\sin\frc{4\pi}\kappa\; \frc{\G\lt(1-\frc4\kappa\rt) \G\lt(-1+\frc8\kappa\rt)}{\G\lt(\frc4\kappa\rt)}~.
\]
With appropriate normalisation, it can be written
\[
	M_N^{(1)} = \sum_{i=1}^{N} x_i G_{{\cal C}^{(i)}}~,\quad
	x_i = \sum_{j=0}^{[N/2]} c_{i,j} A^j B^{[N/2]-j}
\]
where the $c_{i,j}$'s are the unique solution to
\[
	c_{i,j+1} + c_{i+1,j} + c_{i+2,j+1} = 0 ~(i=1,2,\ldots,N-1,\,j=-1,0,\ldots,[N/2])
\]
with $c_{i+N,j} = c_{i,j}$ and
\[
	c_{1,[N/2]}=1,\; c_{2,[N/2]} = \cdots = c_{N,[N/2]} = 0,~ c_{i,-1}=0~.
\]
From the formulas (\ref{bcbosgeneral}) and (\ref{bcgeneral}), one can check that this linear combination
has purely fermionic behavior as $\theta_i\to\theta_{i+1}$ for $i=2,3,\ldots,N$ (with $\theta_{N+1}\equiv \theta_1-2\pi$), and
a bosonic component as $\theta_1\to\theta_2$ (along with some fermionic component).
We have for instance
\[
	M_3^{(1)} = (A+B) G_{{\cal C}^{(1)}} - BG_{{\cal C}^{(2)}} - B G_{{\cal C}^{(3)}}~.
\]

As we said above, it is expected that this solution is, for every $N$, the unique one (up to normalisation)
with those behaviors -- without the need to specify the fermionic component as $\theta_1\to\theta_2$. The requirement for this solution to be a measure is that it be everywhere positive (with appropriate normalisation). We see for instance that $M_3^{(1)} = (A-B) G_{{\cal C}^{(1)}}$ when all angles are equidistant, and that as $\theta_1\to\theta_2$, we have $M_3^{(1)} \sim (A-B)(A+2B) G_{N-2}^f \lt(\sin\frc{\theta_1-\theta_2}2\rt)^{-2r_b}$; at both of these particular points the function $M_3^{(1)}$ has the same sign. It is also possible to check numerically that everywhere it has the same sign. Hence it correctly represents a measure. We also verified that in the case $N=4$ the sign is the same at the particular points where all angles are equidistant and where $\theta_1\to \theta_2$. A general proof of positivity would be very interesting and would strengthen the conjecture according to which the linear combinations above are measures in the $O(n)$ model, but it is beyond the scope of this paper.

When more then one pair is taken, our solutions are not enough to
form linear combinations with the suggested behaviors. One needs
to take certain analytic continuations, which are not obviously
real and positive.

\ack

We thank Denis Bernard for useful discussions and explanations.
This work was supported by EPSRC under grants GR/R83712/01 (J.C.) and GR/S91086/01 (B.D., post-doctoral fellowship).

\appendix

\sect{A short definition of Schramm-Loewner Evolution (SLE)}
\label{defSLE}

Radial SLE (which is the type of SLE of interest for our present work) is a way of constructing a measure $\mu(\gamma)$ for a random (non-self-crossing, continuous) curve $\gamma$ on the unit disk $\mathbb{D}$ joining a point $a\in \partial \mathbb{D}$ of its boundary to the center of the disk, such that a certain property of ``local conformal invariance'' holds. This property is mathematically known as domain Markov property, and says that
\[
	\mu|_{\Gamma \subset \gamma} = \mu \cdot f_\Gamma
\]
where $\Gamma$ is a curve with one end at the point $a$ and the other inside the disk, and $f_\Gamma$ is the {\em uniformizing conformal map} for $\Gamma$, a conformal map $f_\Gamma : \mathbb{D}\setminus \Gamma \to \mathbb{D}$ that maps the disk from which the ``slit'' $\Gamma$ has been removed back to the disk itself, preserving the center (this map is defined up to a rotation). In the equation above, on the left-hand side one restricts the measure to curves $\gamma$ that cover entirely $\Gamma$.

SLE is a construction of the measure $\mu$ through the stochastic growth of a curve from the point $a$ to the center. In general, the growth of a curve $\gamma_t$ with ``time'' $t\in \mathbb{R}^+$ can be described by the growth of its uniformising conformal map $g_t$. The theory of Loewner says that with the uniformising conformal map chosen to have real and positive derivative at the center and with the parametrisation of $t$ given by $g_t'(0) = e^t$, it must satisfy the differential equation
\[
	\frc{\p}{\p t}g_t(z) = -g_t(z) \frc{g_t(z) + a_t}{g_t(z) - a_t}
\]
where $a_t\in\p\mathbb{D}$ is a continuous function from $\mathbb{R}^+$ to the boundary of the disk. This driving function characterises the growing curve that corresponds to the evolving conformal map $g_t$. When the curve is grown to $t\to\infty$, it connects the point $a=a_0$ to the center of the disk. For the grown curve to be a random curve satisfying the property of conformal invariance above, Schramm \cite{Schramm99} found that the random driving function must be a Brownian motion on the boundary of the disk:
\[
	a_t = e^{i\theta_t}~,\quad \theta_t = \sqrt{\kappa} B_t + \theta_0
\]
where $B_t$ is a standard one-dimensional Brownian motion, with normalisation $\mathbb{E} B_t^2 = t$. This describes a one-parameter family of measures, parametrised by $\kappa \in [0,8]$, that satisfy the property of conformal invariance above; these are the only measures with this property.

The power of SLE comes from the fact that probabilities can be evaluated using the explicit growth process of the uniformizing map $g_t$. This generically gives rise to second order linear differential equations which are of the form of level-2 null vector equations of CFT (see, for instance, the review \cite{Cardy05}).

\sect{Derivation of the boundary level-2 null-vector equations on the disk}
\label{derivDj}

The covariance of the correlation function (\ref{corrfct}) under the transformation $z\mapsto z + \alpha(z)$
with (\ref{alpha}) is found by inserting the appropriate charge:
\[
    \lt\bra \phi(e^{i\theta_1})\cdots \phi(e^{i\theta_N})\, \lt(
    \int_C T(z)\alpha(z) \frc{dz}{2\pi i} -
    \int_C \b{T}(\b{z})\overline{\alpha(z)} \frc{d\b{z}}{2\pi i}
    \rt)\,
    \Phi(0)\rt\ket
\]
where $C$ is a contour inside the disk $|z|<1$ going round the
origin counterclockwise once. Using the holomorphic OPE
\[
    T(z) \Phi(0) \sim \frc{h_{\Phi}}{z^2}\Phi(0) +
    \frc{1}z\p\Phi(0) + \ldots
\]
and shrinking the contour $C$ to the origin, we then get
\beq
    (h_\Phi + \b{h}_\Phi) \lt(\sum_{j=1}^N b_j\rt) G~.
\eeq
On the other hand, from the conformal boundary condition for a theory on the disk, the anti-holomorphic
component $\b{T}(\b{z})$ of the stress-energy tensor inside the disk is related to the continuation of the
holomorphic component outside the disk via
\beq\label{confbc}
    \b{T}(\b{z}) = \b{z}^{-4} T(\b{z}^{-1})~.
\eeq
Along with the relation (\ref{presD}), deforming the countour towards the boundary of the disk gives
\[\ba{l}
    \int_C T(z)\alpha(z) \frc{dz}{2\pi i} -
    \int_C \b{T}(\b{z})\overline{\alpha(z)} \frc{d\b{z}}{2\pi i}
    =\z
    \int_C T(z)\alpha(z) \frc{dz}{2\pi i} -
    \int_{C'}
    T(z)
    \alpha_j(z) \frc{dz}{2\pi i} \ea
\]
where the contour $C'$ is outside the disk $|z|<1$ and going counterclockwise. Hence, we are left with
\beq\label{integb}
    -\oint_{z_1,\ldots,z_N}
    \lt\bra T(z) \phi(z_1)\cdots \phi(z_N)
    \Phi(0)\rt\ket\,\alpha(z)\,\frc{dz}{2\pi i}
\eeq
where the integral means a sum of integrals counterclockwise around the points $z_1,\ldots,z_N$.

The relation (\ref{confbc}) specialised to the boundary $z = z_B$ with $|z_B|=1$
implies a set of relations (a Virasoro algebra isomorphism preserving primary fields)
\beq
    \b{L}_n = (-1)^n \b{z}_B^{2n} \sum_{k\ge0} \b{z}_B^k
    \frc{(2-n-k)_k}{k!} L_{n+k}~z
\eeq
amongst the modes $\b{L}_n$ and $L_n$ of the stress energy tensor,
\[
    T(z) = \sum_{n\in\Z} (z-z_B)^{-n-2} L_{n}~,\quad
    \b{T}(\b{z}) = \sum_{n\in\Z} (\b{z}-\b{z}_B)^{-n-2} \b{L}_{n}~,
\]
when they are applied on a boundary field $\phi$ at $z_B$. Along with the Ward identity associated to rotations,
\[
    (z_B L_{-1} - \b{z}_B\b{L}_{-1})\phi(z_B) = (z_B\p\phi-\b{z}_B\b\p\phi)(z_B)
\]
where $\p \equiv \p/\p z$ and $\b\p \equiv \p/\p\b{z}$, we then have for a primary boundary field
\beq\label{actionL}\ba{l}
    L_n\phi(z_B) = 0 \quad (n\ge1) ~,\quad L_0\phi(z_B) = h\phi(z_B)~,\z
    L_{-1}\phi(z_B) = [z^{h} \p (z^{-h}\,\phi(z))]_{z=z_B}~.\ea
\eeq
For a level-2 degenerate boundary field, on which $L_{-2} = \frc{\kappa}4 L_{-1}^2$, this gives
\beq\label{actionL2}
    L_{-2}\phi(z_B) = \frc{\kappa}4 [z^{h} \p^2 (z^{-h}\,\phi(z))]_{z=z_B}~.
\eeq
This can be used to evaluate (\ref{integb}), giving
\[
    \sum_{j=1}^N b_j\,z_j^h\t{\cal D}_j z_j^{-h} G
\]
with
\beqa &&
    \t{\cal D}_j = -\frc{\kappa}2 \lt(\frc{\p}{\p \theta_j}\rt)^2
    + \lt(\frc{\kappa}2-3\rt)i\frc{\p}{\p\theta_j} + \frc{6-\kappa}{2\kappa} \n
    && \quad - \sum_{k\neq
    j}\lt(\cot\lt(\frc{\theta_k-\theta_j}2\rt)\frc{\p}{\p\theta_k}
    + ih\,\cot\lt(\frc{\theta_k-\theta_j}2\rt)
    - \frc{h}{2\sin^2\lt(\frc{\theta_k-\theta_j}2\rt)}\rt)~.\no
\eeqa
Using the similarity transformation
\[
    z_j^h \frc{\p}{\p\theta_j} z_j^{-h} = \frc{\p}{\p\theta_j} - ih~,
\]
we finally find the null-vector equations for the correlation function $G$ to be (\ref{nveq}) with
differential operators (\ref{Dj}).

\sect{Derivation of the constraints from null-vector equations}

\label{app2}

The equation ${\cal D}_1 G = (d_{\Phi_1^f}+\Delta) G$ leads to two equations, upon equating the coefficients of
$\theta_{1,2}^{-2r-1}$ and $\theta_{1,2}^{-2r}$:
\beq\label{1.1}
    \p_2 A - \frc{(r \kappa-3)(2r\kappa-\kappa+2)}{2\kappa}\,B =0
\eeq
and (\ref{1.2}).
On the other hand, the equation ${\cal D}_2 G = (d_{\Phi_1^f} + \Delta) G$ leads to two similar-looking but different
equations:
\beq\label{2.1}
    r\kappa\p_2 A + \frc{(r \kappa-3)(2r\kappa-\kappa+2)}{2\kappa}\,B =0
\eeq
and
\beq\label{2.2}\ba{l}
    \sum_{k\neq1,2} ((f_{2k}\p_k -hf_{2k}')A - \frc16
    (2r-h)A - \frc{\kappa}2 \p_2^2 A - \kappa(2r-1)\p_2 B
    -\z - \frc{(2r\kappa-\kappa-6)(r \kappa-\kappa+1)}\kappa\, C = \Delta A~.\ea
\eeq
Equations (\ref{1.1}) and (\ref{2.1}) imply (\ref{cond1}).
On the other hand, it is a simple matter to check that these conditions automatically lead to the
consistency of equations (\ref{1.2}) and (\ref{2.2})

Thirdly the equations ${\cal D}_j G = (d_{\Phi_1^f}+\Delta) G$ for $j\ge 3$ lead to
\beq\label{3.1}
    -\frc{\kappa}2 \p_j^2A + \sum_{k\neq1,2,j} (f_{jk}\p_{k} - hf_{jk}')A
     + f_{j2}\p_{2} A - 2(h-r)f_{j2}'A = \Delta A~.
\eeq
In the bosonic case $r=r_b=h$, this along with condition (\ref{cond1}) simply gives (\ref{3.2}),
and in the fermionic case $r=r_f$, we find (\ref{3.3}).

\sect{The case $N=2$}
\label{N2}

From the viewpoint of the Calogero-Suthgerland hamiltonian, the
case $N=2$ is not of great interest. Indeed, since the
eigenfunctions just depend on the single variable
$\theta_1-\theta_2$, it is a simple matter to obtain a general
solution to the Calogero-Sutherland eigenvalue equation. Allowing
arbitrary boundary conditions both as $\theta_1\to\theta_2^+$ and
as $\theta_1\to(\theta_2+2\pi)^-$, any eigenvalue can be obtained
(we do not discuss issues associated to the Hermiticity of the
hamiltonian in such conditions). Moreover, the two null-vector
equations are equivalent, hence such a general solution satisfies
all required properties of conformal correlation functions. Any
bulk field dimension $d_\Phi$ is then allowed to appear. However,
of course, not all are expected to correspond to dimensions of
actual fields of the underlying CFT. It is then instructive to
enumerate and interpret some scaling dimensions associated to
known fields.

Besides the 2-leg exponent discussed above,
there are three scaling dimensions known to correspond to well-defined $O(n)$ configurations that we
wish to discuss.

One is the dimension 0, corresponding to the indicator event: it is associated to the measure on curves started
at some angle $\theta_1$ and ended at $\theta_2$ that enclose the origin. Of course, no ``shrinking'' disk
around the origin is involved in the definition of this measure, hence the associated exponent is trivially 0.
The corresponding appropriately normalised correlation function
gives Schramm's formula \cite{Schramm99} (on the disk), derived in the context of SLE:
\[
    G_2^{Schramm} = e^{i\frc{\theta}2} \sin^{\frc2{\kappa}} \lt(\frc\theta2\rt)
    {\ }_2F_1\lt(1,\frc4\kappa;\frc8\kappa;
    1-e^{i\theta}\rt)~,\quad \theta=\theta_1-\theta_2~.
\]
By definition, at $\theta=0$ the hypergeometric function is on its principal branch.
It has a branch point at its argument equal to 1, hence as $\theta$ goes from 0 to
$2\pi$, the branch point is circled once counterclockwise and a monodromy is acquired.

\begin{figure}[t]
\bc \includegraphics{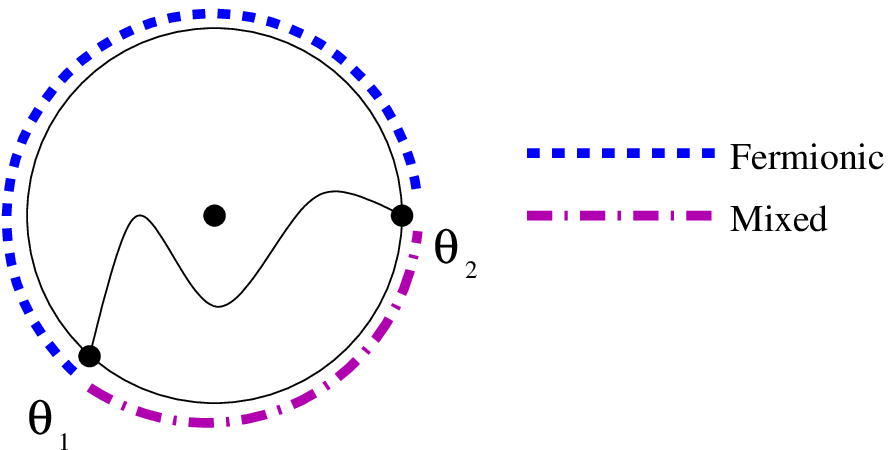} \ec
\caption{A typical configuration for the measure associated to the indicator event, with
boundary conditions.}
\label{figIndic}
\end{figure}

The corresponding Calogero-Sutherland eigenfunction has purely fermionic boundary condition
on the side where the SLE curve surrounds the origin ($\theta\to0^+$),
and a mix of bosonic and fermionic on the other side ($\theta\to 2\pi^-$),
as depicted in Fig. \ref{figIndic}.
Again, the boundary conditions can be made plausible. On the purely fermionic side,
the bosonic behavior does not occur because it would mean imposing a macroscopic loop (since the loop
has to be near to the boundary and has to surround the origin); such macroscopic loops do not occur
with probability 1 in the measure on loops.
On the other side, the bosonic behavior does occur, because imposing a loop of any size not surrounding the origin
does not affect the measure; they do occur with probability 1.
The fermionic fusion also occurs on both sides since the
two curves starting at one point are allowed to extend macroscopically in both cases.

Another is the dimension
\beq
    d^m_2 = \frc{(6-\kappa)(\kappa-2)}{8\kappa} + \frc{\kappa^2-16}{32\kappa}
\eeq (corresponding to the eigenvalue $E_2^{m} = \frc{1}{16}$ of
the Calogero-Sutherland hamiltonian). It is natural to consider
this dimension, since the associated correlation function, \beq
    G^{1-arm}_2 =
    \sin^{-2r_f}\lt(\frc{\theta_1-\theta_2}4\rt)\cos^{-2r_b}\lt(\frc{\theta_1-\theta_2}4\rt)
\eeq
gives purely fermionic boundary condition on one side, and purely bosonic on the other side.

As was noted in \cite{Cardy04}, when
specialised to $\kappa=6$, it corresponds to the one-arm exponent\footnote{
The one-arm exponent characterises the power law with which vanishes the measure
in site percolation with constraint that at least one path exists from the origin to a surrounding circle,
as the radius of the circle is made infinite}, calculated in the context of SLE in
\cite{LawlerSW02}. More generally, for $\kappa>4$ it gives $h/6+c/12+\lambda$ where $\lambda$ occurs
in the measure $\mu\propto \varepsilon^\lambda$ as $\varepsilon\to 0$
on single radial SLE curves that contain no counter-clockwise loops
around the origin before reaching a radius $\varepsilon$ to the origin \cite{LawlerSW02}.
Hence, we expect that the quantity $G^{1-arm}_2\, \varep^{d^m_2}$
give the leading part of the measure on single curves connecting
points at angles $\theta_1$ and $\theta_2$ on the boundary of the disk with the condition that no
loop forms around the origin unless it is completely contained inside a disk of radius $\varep$ around the origin.
The extra terms $h/6+c/12$ in $d^m_2$ account for the change from a radial curve (starting on the boundary
and ending at the center) to chordal curve (starting and ending on the boundary).

This interpretation is corrobated by noticing that the dimension (\ref{dimnprime})
of the fields $\Orr_0$ in the continuum $O(n)$ model is exactly the exponent $d_2^m$. Recall that the field
$\Orr_0$ placed at the origin forbids any loop surrounding the origin in the $O(n)$ model, since it attributes to them
a weight 0. Naturally, the dimension of this field at $\kappa=6$ is the one-arm exponent, since
the absence of loops around the origin implies the presence of a percolation path from the boundary of the disk
to the center.

From this interpretation, the boundary conditions
can be understood as follows. On the side of the bosonic behavior, the fermionic fusion is absent because
two curve starting from one point will almost surely, for $\kappa>4$, have double points so that loops are formed
around the origin; if this is forbidden, the two curve cannot extend macroscopically and the fusion does not occur.
On the side of the fermionic behavior,
the bosonic behavior is absent because joining the curve exactly produces a forbidden loop around the origin.

The last dimension that we wish to consider here is the 0-leg exponent
((\ref{Nlegexp}) with $N=0$), which turns out to be $d^f_0=(c-1)/12$.
The corresponding correlation function can be read off from our solution with $N=2$ and $R=1$,
\beq
    G^{0-leg}_2 = e^{-i\frc{(\kappa+4)\theta}{2\kappa}} \sin^{1-\frc6{\kappa}} \lt(\frc\theta2\rt)
    {\ }_2F_1\lt(1-\frc4\kappa,1-\frc4\kappa;2;
    1-e^{i\theta}\rt)
\eeq
with $\theta=\theta_1-\theta_2$ (this corresponds to the eigenvalue $E_2^{0-leg} = 0$ of the
Calogero-Sutherland hamiltonian). It gives purely bosonic
conditions on one side ($\theta\to 0^+$), and mixed on the other
side ($\theta\to 2\pi^-$). We expect that the quantity
$G^{0-leg}_2\,\varep^{d^f_0}$ is the leading part of the measure
on configurations where a curve joins points at angles $\theta_1$
and $\theta_2$ on the boundary while being restricted not to come
closer than $\varep$ to the origin. It is natural that this
amplitude diverge as the radius is sent to zero (the 0-leg
exponent is indeed negative, except when $\kappa=4$, where it is
0). This interpretation is re-inforced by noticing the following.
It is a simple matter to observe that for the maximum value $n'=2$
(that is, $\kappa'=4$), one finds that the field $\Orr_{n'}$ of the
continuum $O(n)$ model has a dimension (\ref{dimnprime}) given by
the 0-leg exponent $d_{2,n} = d^f_0$. For this maximum value,
there is more likely a loop around the origin, which constrains
the curve to stay away from the origin as described above. The
boundary conditions can also be understood from this picture. On
the bosonic side, there is no fermionic fusion because the two
curves starting from one point are restrained away from the origin
(they cannot form small enough loops around the origin). On the
mixed side, fermionic contributions are clearly non-zero, and the
bosonic fusion occurs since adding a macroscopic loop around the
origin does not change the measure (such loops are already very
likely).

Note that it is this situation that we generalised to $N$ particles with a $N'$-leg bulk field, $N'=N-2M,\,M\in \N$.

\end{document}